\documentclass[12pt]{iopart}
\usepackage{epsfig,graphicx, pictex}
\usepackage{iopams}
\newtheorem{theorem}{Theorem}

\newcommand{\imp}{\:\Rightarrow\:}

\begin{document}

% \noindent version of \today
\noindent version of  28 March 2013

 \title[WKB Approximation to the Power Wall]
 {WKB Approximation to the Power Wall}

 \author[F Mera, S Fulling, J Bouas, and K Thapa]
 {F D Mera$^1$, S A Fulling$^{1,2}$, J D Bouas$^1$, and K Thapa$^2$}

 \address{$^1$ Department of Mathematics, Texas A\&M University,
 College Station, TX, 77843-3368 USA} 
 
 \address{$^2$ Department of  Physics, Texas A\&M University,
 College Station, TX, 77843-4242 USA} 

\eads{\mailto{merandi12@gmail.com}, \mailto{fulling@math.tamu.edu},
\mailto{jd.bouas@gmail.com}, \mailto{thapakrish@gmail.com}}

\begin{abstract}
 We present a semiclassical analysis of the quantum propagator of a 
particle confined on one side by a steeply, monotonically rising 
potential.
 The models studied in detail have potentials proportional to 
 $x^{\alpha}$ for $x>0$;
 the limit $\alpha\to\infty$ would reproduce a perfectly reflecting 
boundary, but at present we concentrate on the  cases $\alpha =1$ 
and~$2$, for which exact solutions in terms of well known functions 
are available for comparison.
We classify the classical paths in this system  by their 
qualitative nature and calculate
the contributions of the various classes 
 to the leading-order semiclassical approximation:
 For each classical path we find the action $S$, 
the amplitude function $A$ and the Laplacian of $A$. (The 
Laplacian is of interest because it gives an estimate of the error 
in the approximation 
 and is needed for computing higher-order approximations.)
The resulting
  semiclassical propagator can be used to rewrite the exact problem 
  as a Volterra integral equation, whose formal solution by iteration
(Neumann series) is a semiclassical, not perturbative, expansion.
 We thereby test, in the context of a 
concrete problem, the validity of the two technical hypotheses 
in a previous proof of the convergence of such a Neumann series in 
the more abstract setting of an arbitrary smooth potential. 
 Not surprisingly, we find that the hypotheses are violated when 
caustics develop in the classical dynamics; this opens up the 
interesting future project of extending the methods to momentum 
space.
\end{abstract}

 \pacs{03.65.Sq, 02.30.Rz}
 \ams{81Q20, 45D05}
 \submitto{\JPA}

 \maketitle

 \section{Introduction} \label{intro}

This article continues the semiclassical 
analysis that was started in \cite{fuldar} and \cite{Mera}. 

In \cite{Mera,Mera2}, a general theorem about the construction 
of solutions of Volterra integral equations was proved;
\cite{Mera} also outlined an application to semiclassical 
approximation for the time-dependent Schr\"odinger equation
(converted to an equivalent integral equation).
The main idea \cite{BB74,dowkw}
% due to Balian and Bloch \cite{BB74}, 
is to use the WKB approximation to the quantum 
Green function (propagator) as the foundation for the 
solution of the integral equation by iteration.
In \cite[Theorem 12]{Mera},  which we call the 
``semiclassical Volterra theorem''\negthinspace,
it was proved that this series solution will converge under 
two technical hypotheses.  However, the issue of whether 
those hypotheses are satisfied in particular concrete problems 
was not studied there.

In the present paper we apply the semiclassical Volterra theorem 
to the family of potentials $z^{\alpha}\theta(z)$ in $\mathbb R^1$ and
test the validity of the  hypotheses.  This study was begun in 
\cite{fuldar}, which also introduced the corresponding 
potentials in $\mathbb R^n$ as a model of boundary effects on the 
vacuum energy in quantum field theory.
Here we consider in detail the
simplest two cases, $\alpha=2$ and $\alpha=1$.

The semiclassical construction is based on the classical 
paths (of the dynamical system in question)
between two space-time points.
 Therefore, the first step is the classification of 
these trajectories.  For example, if the starting and ending 
point are both in the potential-free region (left of the vertical 
axis), there is always a direct path between them that never 
enters the potential; its contribution to the semiclassical 
propagator is just the free quantum  propagator.
For the potentials under study, there is usually one other 
 zeroth-order path, which enters the potential region and 
bounces back.  We find that this path
 goes through a 
caustic (focal point), so that the WKB approximation to that 
path's contribution to the quantum propagator must be modified 
by a Maslov index.
Similarly, we analyze the possible paths when one or both of the 
points is inside the potential.

 For each classical path we construct the action and 
amplitude functions that define the contribution of that path to 
the semiclassical approximation to the propagator. The exact 
propagator is a solution of the time-dependent Schr\"odinger 
equation; applying the Schr\"odinger differential operator to any 
of the semiclassical terms leaves a residual expression 
proportional to the Laplacian of the amplitude function.
 The proof of the semiclassical Volterra theorem shows how to pass 
from the residual to an estimate on the error of a semiclassical 
solution.  Furthermore, when the Laplacian function is bounded
  (no caustics are encountered in the region concerned), a 
first-order (in~$\hbar$) correction can be calculated by 
concatenating two classical paths and integrating a certain 
integrand over the location of the point where they are joined.
 If the potential is smooth. this process can be continued to 
arbitrarily high order, although the formulas rapidly become very 
cumbersome, and the theorem shows that this series in $\hbar$ is 
convergent to the exact solution.

 In the model under study this idealized strategy is eventually 
disrupted in two ways. 
 First, our model potential has a mild singularity at $z=0$, so the 
semiclassical expansion fails at a sufficiently high order.
Second, and more fundamentally, caustics do occur, even where the 
potential is smooth, and they cause blowups in the residuals, error 
estimates, and higher-order corrections.  The ultimate remedy for 
this disease is to convert the calculations to momentum space
 \cite{MF,Little,Todd}, 
 a project that goes far beyond the scope of the present paper.

    \Sref{theory} reviews general semiclassical (or WKB) theory, 
    and  \sref{pwall}  sets up a class of models, the power walls,
  and begins the analysis of their classical paths.
 Detailed semiclassical treatments of quadratic and linear power 
walls are presented in \sref{quad} and \sref{linear}, 
 respectively, and \sref{concl} presents some conclusions.

\section{Hamilton--Jacobi Theory, WKB Approximation
   and Volterra Integral Equation} 
\label{theory} 

 In this section we review the general theory of construction of 
the semiclassical propagator
  $K_\mathrm{scl}(\mathbf{x},t;\mathbf{y},s)$, following such 
  treatises as \cite{BB97} and~\cite{MF}.

Consider  a quantum particle subject to a (sufficiently smooth)
potential $V(\mathbf{x},t)$, $\mathbf{x}\in\mathbb{R}^n$.
 A natural ansatz for the wave function is
\begin{equation} \label{wavefn} 
\psi(\mathbf{x},t) = A(\mathbf{x},t) e^{\frac{i}{\hbar} 
S(\mathbf{x},t)} \end{equation}
 where $A(\mathbf{x},t)$ and 
$S(\mathbf{x},t)$ are called the amplitude and the action of 
$\psi(\mathbf{x},t)$, respectively.  Substituting
\eref{wavefn} into the time-dependent Schr\"{o}dinger equation
 \begin{equation}\label{schrod}
 i\hbar\,{\partial \psi\over \partial t} = -\,{\hbar^2\over 
2m}\nabla^2\psi + V\psi , \end{equation}
 one obtains the  partial differential equation
\begin{equation}
\fl{
0 = A \biggl [ \frac{\partial S}{\partial t} + 
\frac{1 }{2m} (\nabla S)^{2} + V \biggr ] - i \hbar \biggl 
[\frac{\partial A}{\partial t } + \frac{1}{m}(\nabla A \cdot 
\nabla S)+ \frac{1}{2m} A \Delta S \biggr ] - 
\frac{\hbar^{2}}{2m} \Delta A,
}
\label{SAwaveeq}\end{equation}
where $ \Delta $ is the Laplacian operator.
  Separate the real and imaginary parts 
of \eref{SAwaveeq} to get 
\begin{equation} 
\label{HJquantum} \frac{\partial S}{\partial t} + \frac{1}{2m} 
(\nabla S)^{2} + V = \frac{\hbar^{2}}{2m} \frac{\Delta A}{A} 
\end{equation} 
and
\begin{equation} \label{transport} m 
\frac{\partial A}{\partial t } + (\nabla A \cdot \nabla S)+ 
\frac{1}{2} A \Delta S = 0 .
\end{equation} 
The classical limit is obtained by taking the limit $\hbar \to 0$, 
 whereupon \eref{HJquantum} becomes 
\begin{equation}\label{HJ} 
 \frac{\partial S}{\partial t} + \frac{1}{2m} (\nabla S)^{2} + 
V(\mathbf{x},t) 
 =0. \end{equation} 
This is the Hamilton--Jacobi 
equation.  The phase $S(\mathbf{x},t)$ is 
interpreted as the classical action.

 Equation \eref{HJ} has the form
 $ \frac{\partial S }{\partial t} + H(x, \nabla 
S(x,t),t)  = 0$,
 where $H$ is the classical Hamiltonian 
function, 
$H(\mathbf{x},\mathbf{p},t) = \frac{1}{2m}| \mathbf{p}|^{2} + 
V(\mathbf{x},t)$.         
 A classical solution is a local 
curve $\mathbf{x}(t)$  satisfying the equations 
\begin{equation}\label{classicaleq} 
\frac{d \mathbf{x}(t)}{d t} = 
\frac{\partial H}{\partial \mathbf{p} } = {\mathbf{p}\over m}
 = \frac{1}{m} \nabla S(\mathbf{x}(t),t) 
\end{equation} 
and
\begin{equation}
\label{classicaleq2} \frac{d \mathbf{p}(t)}{d t} 
= - \,\frac{\partial H}{\partial \mathbf{x} }\,. \end{equation}
 Equation \eref{classicaleq} enables one to 
construct the action $S(\mathbf{x},\mathbf{y},t)$ from a 
knowledge of a family of classical solutions $\mathbf{x}(t)$.
   The total time derivative of the action must be 
\begin{equation} 
\frac{d 
S}{d t} = \frac{\partial S}{\partial t} + \dot{\mathbf{x}} \cdot 
\nabla S = - H + \dot{\mathbf{x} } \cdot p \equiv 
L(\mathbf{x}(t), \dot{\mathbf{x}}(t) ) .
\end{equation} 
This equation implies that we can get solutions of the
  Hamilton--Jacobi 
equation by integrating the Lagrangian $ L $ along the 
trajectories:
%  Thus, the action $ S $ can be defined by 
\begin{equation} 
S(x, y, t) = \int_{0}^{t} L(x(u), \dot{x}(u) ) 
\, du + S_{0}\,, 
\end{equation}
where $ S_{0} $ is initial data, and 
 $ S(x,y, t) $ then solves the Hamilton--Jacobi equation.

 Conversely, 
if we have a local solution of the Hamilton--Jacobi equation, 
 then $\nabla S = \mathbf{p} = m \dot\mathbf{x}$
 along its classical trajectories.
Then to solve the 
Schr\"{o}dinger equation through order $ \hbar $ we still
  need to solve the transport equation \eref{transport}.
 The latter can be rewritten as
\begin{equation}
 - \frac{1}{2m} A \Delta S 
= \biggl ( \frac{\partial }{\partial t} + \frac{1}{m} \nabla S 
\cdot \nabla \biggr ) A = \biggl ( \frac{\partial }{\partial t} + 
\dot\mathbf{x} \cdot \nabla \biggr ) A = \frac{d A}{d t} \,,
\end{equation} 
or
 \begin{equation}
  - \frac{1}{2m} \Delta S = \frac{1}{A} 
\frac{d A}{d t} = \frac{d}{dt} \ln A.
\end{equation}
We can 
solve for $ \ln A $ by integrating along the classical 
trajectories, with the result
\begin{equation} 
A(\mathbf x,t) \propto \exp \biggl [ - \frac{1}{2m } 
\int_{0}^{t} \Delta S(\mathbf x(u), u ) \, du \biggr ]
\end{equation} 
 (where the $\mathbf y$ dependence has been suppressed).
However, the amplitude function can be expressed in an 
alternative way: 
\begin{equation} \label{vvleck}
A(\mathbf x,t)
 \propto
\sqrt{\det C}, 
\end{equation} 
where $ C = \nabla_{x} \nabla_{y} S\, $;
  $ \det C $ is known as the Van Vleck determinant. 
 The fact that this determinant is a solution of the 
 transport equation is well known but 
 nontrivial~\cite{BB97}.  

In what follows we  restrict attention to time-independent 
 potentials $V(\mathbf x)$ and concentrate on the 
 quantum propagator, the Green function that produces a solution of 
the time-dependent Schr\"odinger equation from arbitrary initial 
data. 
  $K$~can  be written in Dirac notation  as 
\begin{equation}  \label{Umatrix}
  K(\mathbf x,t;\mathbf y,s) =
  \langle \mathbf x | \hat{U}(t,s) | \mathbf y \rangle,
\end{equation} 
where $\hat{U}(t,s)$ is the unitary 
time-evolution operator for the system taking states at time $s$ 
to states at time $t\,$. 
 Intuitively, the  propagator $K(\mathbf x,t;\mathbf y,s)$ describes 
the motion of a quantum-mechanical particle travelling from the 
space-time point $(\mathbf y,s)$ to the point $(\mathbf x,t)$ 
 and can be interpreted as 
passing through each possible intermediate point $(\mathbf r,\tau)$ 
 with a certain probability amplitude.
 The basic concept 
that underlies the theory of higher-order semiclassical 
approximations
is that 
in a local space-time region the particle evolves under the 
semiclassical propagator between encounters with the effective 
potential $V_\mathbf{scl}\equiv \frac{\Delta A }{ A}$. 
 (This idea was developed by Balian and Bloch \cite{BB74} in the 
context of the time-independent Schr\"odinger equation and its 
Green function, and mentioned by 
Dowker \cite[sec.~3.2]{dowkw} for the time-dependent case.)
 
  For time-independent $V$, $K$ is a function of the time difference
 $t-s$, so one usually sets $s=0$ without loss of generality,
 and the fourth argument of $K$ is suppressed in the notation.
 (Similarly, the action and amplitude functions in full generality 
are functions of $s$ as well as of $(\mathbf x,t$ and $\mathbf y$.) 
The Green function   $K(\mathbf x,t; \mathbf y,0)$ 
satisfies the homogeneous Schr\"{o}dinger equation in the variables 
$(\mathbf x,t)$, except at the source point $(\mathbf y,0)$. 
Therefore, the machinery introduced  above applies to it.
 By a standard argument (e.g.,~\cite{HJ}), 
 it can also be defined by the nonhomogeneous equation
 \begin{equation} \label{Kschrod}
 \left( H_x - i\hbar \frac{\partial}{\partial t} \right) 
K(\mathbf x,t;\mathbf y,s) = 
 -i\hbar \delta(\mathbf x-\mathbf y)\delta(t-s), 
 \label{nonhomgreen}\end{equation} 
where $H_{x}$ is the Hamiltonian of the quantum system, appearing 
here as a function of the   $\mathbf x$ variable,
  and $\delta$ is the Dirac delta function. 
 Under suitable technical conditions the solution of the 
nonhomogeneous equation
 \begin{equation}
\left(- H + i\hbar \frac{\partial}{\partial t} \right) 
\psi(\mathbf{x},t) = \phi(\mathbf{x},t)
\label{nonhomschrod} \end{equation}
 with initial data $\psi(\mathbf{x},0)=0$ is
\begin{equation}
\psi(\mathbf{x},t) =\int_0^t \int_{\mathbb{R}^n} 
 K(\mathbf x,t;\mathbf y,\tau) \phi(\mathbf{y},\tau)\, d\mathbf{y}\, 
d \tau \equiv [K\phi](\mathbf{x},t) .
\label{nonhomsol}\end{equation} 
With respect to $t$, the integral operator in \eref{nonhomsol}
 is of the Volterra type;
 that is, the upper limit is the solution's variable~$t$,
 rather than $+\infty$ or some large fixed~$T$.

 The free propagator $K_\mathrm{f} (\mathbf x,t; \mathbf y,s)
\equiv K_\mathrm{f} (\mathbf x,\mathbf y,t-s) $ 
in the space-time $\mathbb R^{n} \times \mathbb{R^{+}}$ with 
$V(\mathbf x,t)=0$
is well known to be  (for $t>s$)
\begin{equation} \label{freepropagator}
 K_\mathrm{f} (\mathbf x,t; \mathbf y,s) = \left( 
\frac{m}{2 \pi i \hbar ( t-s)} \right)^{n/2} 
 e^{im|\mathbf x-\mathbf y|^{2}/2\hbar( t-s)}. 
\end{equation} 
The exponent in $K_{f}(\mathbf x,\mathbf y,t)$ is 
  $ \frac{i}{\hbar}$ times the 
action $S_0(\mathbf x,\mathbf y,t)$  for a free 
particle. The Van Vleck determinant for this case is
\begin{equation} 
\det \biggl (- 
\frac{\partial^{2} S_{0}}{\partial x_{i} \partial y_{j}} \biggr) 
= \biggl ( \frac{m}{t} \biggr )^{n}.
\end{equation} 
Thus the free
quantum  propagator fits into the WKB framework as 
\begin{equation} 
\fl{
K_\mathrm{f} (\mathbf x,\mathbf y,t) 
 = A_{0}(t) e^{\frac{i}{\hbar} S_{0}(\mathbf x,\mathbf y,t)}
  = \biggl (\frac{1}{2 \pi i \hbar} \biggr)^{n/2} \sqrt{\det 
\biggl(-\,\frac{\partial^{2} S_{0}}{\partial x_{i} \partial y_{j} } 
\biggr)} \exp \biggl \{\frac{i}{\hbar} S_{0}(\mathbf x,\mathbf y,t)
  \biggr \} }.
\end{equation}
 The particular normalization factor in \eref{freepropagator} is 
the one that gives the correct 
initial value to $K(\mathbf x,t; \mathbf y,s)$ on the surface $t = s$.  
Alternatively, if one thinks of $K(x,t; y,s)$ as a solution of 
the nonhomogeneous Schr\"{o}dinger equation \eref{Kschrod}
  in all of space-time, it 
gives the correct delta-function singularity at $(x,t)$ = 
$(y,s)$.
  
Similarly, we will be able to write a semiclassical propagator 
$K_\mathrm{scl}(\mathbf  x, t; \mathbf y, s)$ 
 in the form
  \begin{equation} 
K_\mathrm{scl}(\mathbf x,t; \mathbf y,s)
  = (2 \pi i \hbar)^{-n/2} \sqrt{\det C} \,
e^{iS/\hbar} \end{equation}
  where $C$ is the $n \times n$ matrix
with elements  $C_{ij} = - \frac{\partial^{2} S}{\partial 
x_{i} \partial y_{j}}$. The factor $\sqrt{\det C} e^{iS/\hbar} $ 
arises as the solution of the transport equation 
\eref{transport}, and the arguments for the normalization factor
 are the same as in the free case.

A well-known technique \cite{folland,kress,rub}
  for the construction of Green
functions for the Laplace and Helmholtz equations, 
 and also the heat equation, in bounded
domains in $\mathbb R^n$ (billiards) is by reduction to integral
equations on the boundary.  An important feature of the heat 
equation is that the solution of the boundary integral equation 
by iteration is convergent because of  its 
Volterra structure. Therefore one has, in principle, an explicit 
construction of the solution. The Schr\"{o}dinger equation has 
the same Volterra structure, so one expects again to have a 
convergent series solution.
  A general theorem to this effect was proved in
 \cite[Ch.~6]{Mera} with corrections in~\cite{Mera2}. 
The application of this general Volterra theorem in a particular 
context   reduces to 
showing that the operator family arising in the construction of the 
individual terms in the series is uniformly bounded on a suitable 
Banach space. 
In any particular case this may be 
 a nontrivial task and may require 
additional technical assumptions.

 In the semiclassical Schr\"odinger problem \eref{schrod},
  the key idea that we implement
 is to use the WKB approximation to the quantum 
kernel analogously to the free kernel approximation in billiard 
problems. In the billiard problem,  scattering happens only at the 
boundary;  in our case, the particle is scattered throughout the 
bulk region by a source that is the residual error in the WKB 
approximation to the exact kernel. This construction is developed 
in \cite[Ch.~8]{Mera}. The WKB kernel is 
\begin{equation}\label{WKBkernel} K_\mathrm{scl} (\mathbf{x}, t; 
\mathbf{y}, 0) = (2\pi i \hbar )^{-n/2} A e^{iS/ \hbar}, 
\end{equation}
  where, as explained above, 
 \begin{equation}\label{classicalaction} 
S(\mathbf{x},\mathbf{y}, t) = \int_{0}^{t} L(\mathbf{q}(\tau), 
\dot{\mathbf{q} }(\tau)) \, d\tau, \quad L(\mathbf{q}(\tau), 
\dot{\mathbf{q} }(\tau))  = \frac{m}{2} \dot{\mathbf{q} }^2 - 
V(\mathbf{q} ),
  \end{equation} 
 is the classical action, 
 and the amplitude $A$ is 
\begin{equation}\label{amplitude} A(\mathbf{x}, \mathbf{y}, t) = 
\sqrt{\det \biggl | \frac{\partial^2 S}{\partial x_{i} \partial 
x_{j} } \biggr |}\,. 
 \end{equation} 

If there is more than one classical trajectory $\mathbf{q}(\tau)$ 
starting at $\mathbf{y}$ at time $0$ and arriving at $\mathbf{x}$ 
at time $t$, the semiclassical approximation is a sum of such 
terms, possibly modified by Maslov phase factors to keep track of 
places where the radicand in \eref{amplitude} becomes negative. 

 We define a kernel $Q$ by 
\begin{equation}\label{Q} 
Q(\mathbf{x}, t; \mathbf{y}, \tau ) 
 = (2\pi i \hbar )^{-n/2}\hbar^{2} [\Delta_{\mathbf{x} } 
A(\mathbf{x}, t; \mathbf{y}, \tau )] 
e^{iS(\mathbf{x},t;\mathbf{y},\tau )/\hbar}.
 \end{equation}
 Both $Q$ and $K_\mathrm{scl}$ define Volterra operators by 
formulas precisely analogous to~\eref{nonhomsol};
 we denote these operators (including both time and space 
integrations) by the same letters as the kernels.
  The  operators $K$, $Q$ and $K_\mathrm{scl}$ are related by 
\begin{equation}
K^{-1}K_\mathrm{scl} = (- i \hbar \partial_{t} - \hbar^{2} \nabla^{2} + 
V(\mathbf{x}) ) K_\mathrm{scl} = 1- Q \,; 
\end{equation} 
that is, $Q = 
O(\hbar^{2})$ is the amount by which $K_\mathrm{scl}$ fails to 
solve the PDE for which it was devised. 
 Thus, formally 
 \begin{equation}  \label{Qseries}
K= K_\mathrm{scl}(1-Q)^{-1}= K_\mathrm{scl} \sum_{j=0}^{\infty} Q^{j}.
 \end{equation} 
 The Volterra property ought to enable one to prove that this series 
converges.

 To examine this question in detail, we write out the kernel 
representation of~\eref{Qseries}.  We introduce a notation for the 
spatial parts of the integral operators:
 \begin{equation}  
 [\Gamma(t,\tau)\phi](\mathbf{x}) \equiv
[\hat{K}_\mathrm{scl}(t,\tau)\phi](\mathbf{x})
  = \int_{\mathbb R^n} 
K_\mathrm{scl}(\mathbf{x},t;\mathbf{y},\tau) \phi(\mathbf{y},\tau)\, 
d\mathbf{y},
\label{Gammadef}\end{equation}
 \begin{equation}
 [\hat{Q}(t,\tau)\phi](\mathbf{x}) = \int_{\mathbb R^n} 
Q(\mathbf{x},t;\mathbf{y},\tau) \phi(\mathbf{y},\tau)\, 
d\mathbf{y}.
\label{Lambdadef}\end{equation}
 Note that $Q$ from \eref{Q} can be written
\begin{equation}\label{Qfactored} 
Q(\mathbf{x}, t; \mathbf{y}, \tau ) 
 = \hbar^{2}\, {\Delta_{\mathbf{x} } A(\mathbf{x}, t; \mathbf{y}, \tau ) 
 \over A(\mathbf{x}, t; \mathbf{y}, \tau )}
K_\mathrm{scl} (\mathbf{x},t;\mathbf{y},\tau ),
 \end{equation}
 so
 \begin{equation} \label{Lambdafactored}
 [\hat{Q}(t,\tau)\phi](\mathbf{x}) =
  \hbar^{2} \int_{\mathbb R^n} 
{\Delta_{\mathbf{x} } A(\mathbf{x}, t; \mathbf{y}, \tau ) 
 \over A(\mathbf{x}, t; \mathbf{y}, \tau )}
K_\mathrm{scl}(\mathbf{x},t;\mathbf{y},\tau) \phi(\mathbf{y},\tau)\, 
d\mathbf{y}.
 \end{equation}

Let $\psi=K\phi$ and apply~\eref{Qseries}.  The zeroth-order term is
 \begin{equation}
K_\mathrm{scl}\phi(\mathbf{x},t)= 
 \int_0^t [\Gamma(t,\tau)\phi](\mathbf{x})\, d\tau.
 \end{equation}
 Note that if $\Gamma(t,\tau)$ is a bounded operator on 
$L^2(\mathbb{R}^n)$, with bound $C$ independent of $t$ and $\tau$, then
 \begin{equation}
\|K_\mathrm{scl}\phi\|_{L^2(\mathbb{R}^n)}(t) \le C \int_0^t 
\|\phi\|_{L^2(\mathbb{R}^n)} (\tau) \,d\tau
 \le tC \|\phi\|_{L^{\infty,2}(I_t,\mathbb{R}^n)}\,,
\label{zerothbound} \end{equation}
 where the final norm is the supremum of 
$\|\phi\|_{L^2(\mathbb{R}^n)} (\tau)$ over $\tau\in I_t = [0,t]$.
 (For simplicity of notation we consider only $t$ positive.)
 The first nontrivial term in the series is
 \begin{eqnarray} \fl
K_\mathrm{scl}Q\phi(\mathbf{x},t)&= \int_0^t d\tau_1 
\Gamma(t,\tau_1)\left[ \int_0^{\tau_1}d\tau 
[\hat{Q}(\tau_1,\tau)\phi]\right] (\mathbf{x})
 \nonumber\\
 &=\hbar^2   \int_0^t d\tau_1 
\int_{\mathbb{R}^n} d\mathbf{x}_1\,
  K_\mathrm{scl}(\mathbf{x},t,\mathbf{x}_1,\tau_1) 
\times{} \nonumber\\
 &\hphantom{\hbar^2 D}\int_0^{\tau_1}d\tau 
 \int_{\mathbb{R}^n}d\mathbf{y}\, 
 {\Delta_{\mathbf{x}_1 } A(\mathbf{x}_1, \tau_1; \mathbf{y}, \tau ) 
 \over A(\mathbf{x}_1, t; \mathbf{y}, \tau )}
K_\mathrm{scl}(\mathbf{x}_1,t;\mathbf{y},\tau) 
\phi(\mathbf{y},\tau).
\label{firstterm}
  \end{eqnarray}
 Now suppose that 
 \begin{equation}
\left|{\Delta_{\mathbf{x} } A(\mathbf{x}, t; \mathbf{y}, \tau ) 
 \over A(\mathbf{x}, t; \mathbf{y}, \tau )}\right| \le D
 \quad\mbox{for all $\mathbf{x}$, $t$, $\mathbf{y}$, $\tau$}.
\label{Abound} \end{equation}
 Then
 \begin{eqnarray}
|K_\mathrm{scl}Q\phi(\mathbf{x},t)| &\le \hbar^2 D 
\left|  \int_0^t d\tau_1 
\int_{\mathbb{R}^n} d\mathbf{x}_1\,
  K_\mathrm{scl}(\mathbf{x},t,\mathbf{x}_1,\tau_1) 
\times{} \right.\nonumber\\ 
 &\hphantom{\hbar^2 D}
  \left. \int_0^{\tau_1}d\tau \int_{\mathbb{R}^n}d\mathbf{y}\, 
K_\mathrm{scl}(\mathbf{x}_1,t;\mathbf{y},\tau) 
\phi(\mathbf{y},\tau)\right| \nonumber \\
 &=\hbar^2 D   \int_0^t d\tau_1 \int_0^{\tau_1}d\tau 
 [\Gamma(t,\tau_1)\Gamma(t_1,\tau) \phi](\mathbf{x}),
 \nonumber \end{eqnarray}
 and so
 \begin{equation} \fl
 \|K_\mathrm{scl}Q\phi\|_{L^2(\mathbb{R}^n)}(t)
 \le \hbar^2 DC^2 \int_0^t d\tau_1 \int_0^{\tau_1}d\tau\,
 \|\phi\|_{L^2(\mathbb{R}^n}(\tau)
 \le \hbar^2 DC^2  \,\frac{t^2}2\,
 \|\phi\|_{L^{\infty,2}(I_t,\mathbb{R}^n)}\,.
 \label{firstbound} \end{equation}
 Similarly, or by induction, for each $j$
 \begin{equation}
 \|K_\mathrm{scl}Q^{j-1}\phi\|_{L^2(\mathbb{R}^n)}(t)
 \le \hbar^{2j} D^{j-1}C^{j}   
 \,\frac{t^j}{j!}\,\|\phi\|_{L^{\infty,2}(I_t,\mathbb{R}^n)}\,.
 \label{jthbound} \end{equation}
The sum of all such terms is majorized by an exponential series, so 
it converges in the topology of $L^{\infty,2}(I_T,\mathbb{R}^n)$
 for any $\infty\ge T\ge t$. 
 We have thus proved the following.

\begin{theorem}\label{Meratheorem}% {\rm\cite{Mera}}
Suppose that the following two 
hypotheses hold:
 \begin{enumerate}
\item $\displaystyle \left\| \frac{\Delta A}{A} 
\right\|_{L^{\infty}(I_T{}\!^{2}; \mathbb R^{2n})} < \infty$
 \quad(cf.\ \eref{Abound}).
 \item $\hat{K}_\mathrm{scl}(t,\tau)$ (defined in \eref{Gammadef})
 is a uniformly bounded operator from $L^{2}(\mathbb R^n)$ to itself. 
 \end{enumerate}
Then for all $\phi\in L^{\infty,2}(I_T, \mathbb{R}^n)$
the inequalities \eref{jthbound}  hold. 
It follows that the series
 \begin{equation}
K\phi =\sum_{j=0}^\infty  K_\mathrm{scl}  Q^j\phi
\label{Neumannseries1} \end{equation}
converges (for $t\le T$).
In other words, the nonhomogeneous time-dependent Schr\"odinger 
equation \eref{nonhomschrod} can be  solved by iteration
starting from the semiclassical kernel approximation, 
$K_\mathrm{scl}$ (\eref{WKBkernel}, \eref{classicalaction}),
  to the full propagator, $K$ (see~\eref{nonhomsol}). 
\end{theorem}

 This theorem was proved in \cite[Chapter~8]{Mera}.
 The idea of proof is the same as that for a general theorem on the 
solution of Volterra integral equations by iteration, 
 formulated in \cite[Chapter~6]{Mera} with a flawed proof and
successfully proved in \cite{Mera2}.  Because of the complicated 
structure of the expressions $K_\mathrm{scl}Q^j$, it has been more 
convenient to repeat the argument from the beginning rather than to 
force the problem into the mold of the general Volterra theorem.

 Because $Q(\mathbf{x}, t; \mathbf{y}, \tau )$ for fixed $\mathbf{y}$ 
 is not an $L^2$ function of $\mathbf{x}$, we cannot apply the 
argument to get literal convergence of  the series for
 $K(\mathbf{x}, t; \mathbf{y}, \tau ) $ 
 (without the ``smearing function''~$\phi$).  
  Moreover, the convergence is in $L^2$, not pointwise.
 Of course, it is possible that a stronger theorem holds.

 The condition (ii) is expected to hold wherever $K_\mathrm{scl}$ 
is a decent approximation to~$K$ (i.e., away from caustics).  The 
point is that for fixed times the operator defined by $K$ is 
unitary (cf.~\eref{Umatrix}), so that the one defined by 
$K_\mathrm{scl}$ should be approximately unitary and hence bounded. 
 This reasoning would not apply if $L^2$ were replaced by 
$L^\infty$; indeed, the propagators are not bounded functions as 
$t\to0$.

The determinant in \eref{amplitude} is singular at caustics, 
where the mapping from initial velocity data (at $\mathbf y$)
  to $\mathbf x$ 
ceases to be a diffeomorphism. One can expect both conditions
 in Theorem \ref{Meratheorem}
 to become problematic if the orbit goes through a caustic.
A way to go beyond caustics (if necessary) is provided by the
Maslov theory \cite{MF}, as already implemented in a similar
problem in \cite{TT}.

  The construction in Theorem \ref{Meratheorem} 
 implements the Feynman path 
integral idea in a way different from the usual time-slicing 
approach. 
 (A similar observation was made by Putrov~\cite{putrov}
  in a different context.)
 Each term in \eref{Neumannseries1} is an integral over 
classical paths with $j$ scatterings off an effective potential 
 ${\Delta A}/{A}$.

\section{The Power Wall Potential} \label{pwall}
\subsection{The Model}

To test the validity of the two hypotheses in Theorem 
\ref{Meratheorem} in the context of a concrete problem, 
we consider a family of potentials in one dimension, namely
\begin{equation} \label{wallpot}
V(x) = 
\cases{
0   & if  $x  < 0$, \cr
 \lambda x^{\alpha} & if  $x  \geq 0$.
\cr}
\end{equation}
where $1\le\alpha \in \mathbb{R}$.
The study of this ``power wall'' model, in a wave equation, was 
initiated in 
\cite{fuldar} and continued in \cite{okla} in the context of 
quantum vacuum energy, and we hope that our study of the 
associated Schr\"odinger equation will yield new information 
about the spectral density (and hence the vacuum 
energy) of the operator $-\nabla^2 + V$.
(In the vacuum-energy papers there were two transverse 
dimensions, but here we ignore them because their contribution to 
the quantum kernel in dimension 3 is a trivial factor.)
In \cite{fuldar} the coupling constant 
$\lambda$ was written for any $\alpha$ in terms of a 
dimensionless constant 
and a fundamental length, but here we are concerned with 
particular values of $\alpha$ and will choose the physically most 
natural notation in each case.

The most calculationally tractable values of $\alpha$ are 2 and~1.
They are investigated in detail in the next two sections.

 \subsection{Classification of the Classical Paths}

A particle moving in the potential \eref{wallpot} is acted on by 
a force that never points to the right.  Therefore, the possible 
trajectories have only a small number of possible 
``topologies''\negthinspace.  Consider first the initial-value 
problem, 
where $q(0)=y$ and $\dot q(0) =v =2p$ are prescribed and one 
solves for $q(\tau)$.
If $y\le0$, the particle initially moves freely; if also $p\le 0$,
 it will move freely forever, but if $p>0$ it will eventually 
enter the region with the potential.  In the latter case it will 
accelerate to the left and eventually exit from the potential 
and move freely again.
If $y>0$, the particle immediately accelerates to the left and 
eventually reaches the free region.

\begin{figure}
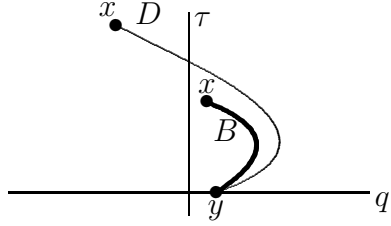

\[
\beginpicture
 \setcoordinatesystem  units <1.2cm, 0.6cm>
 \putrule from -2 0 to 2 0
 \putrule from 0 -0.5 to 0 4
 \put{$\bullet$} at -0.81 3.7
  \put{$\bullet$} at 0.2  2
 \put{$\bullet$} at 0.3  0
 \plot 0 2.9  
       -0.8 3.7 /

 \setquadratic
\setsolid\noindent\plot 0.3 0
 1 1.25
 0 2.9 /  
 \setplotsymbol({\bf.})
\noindent\plot  0.3 0
 0.75 1.1
 0.2 2 /

\put{$x$} [rb] <0pt, 2pt> at 0.3  2.1
\put{$x$} [rb] <0pt, 2pt> at -0.8  3.8
\put{$\tau$} [lt] <2pt, 0pt> at 0 4
\put{$q$} [lt] <2pt, 0pt> at 2 0
\put{$y$} [t] <0pt, -3pt> at 0.3 0
\put{$B$} [rb] <0pt, 4pt> at 0.55  0.9
\put{$D$} [rb] <0pt, 4pt> at -0.3 3.5
\endpicture
\]
 \caption{Classification of  paths with $y>0$.
Solid: Type D. Heavy: Type B.}
 \label{orbitsI}
 \end{figure}

\begin{figure}
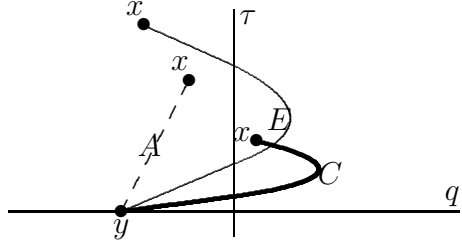

\[
\beginpicture
 \setcoordinatesystem  units <1.5cm, 0.67cm>
 \putrule from -2 0 to 2 0
 \putrule from 0 -0.5 to 0 4
 \put{$\bullet$} at -1 0
  \put{$\bullet$} at -0.8  3.7
 \put{$\bullet$} at 0.2 1.4
  \put{$\bullet$} at -0.4 2.6
 \plot -1 0
  0 0.95 /
 \plot 0 2.9  
       -0.8 3.7 /
 \setdashes\noindent\plot -1 0 
  -0.4 2.6 /
 \setquadratic
\setsolid\noindent\plot 0 0.95
 0.5 1.85
 0 2.9 /  
 \setplotsymbol({\bf.})
  \noindent\plot  0 0.3
 0.75 0.8
 0.2 1.4 /
  \setlinear\noindent\plot -1 0
       0 0.3 /
\put{$q$} [rb] <0pt, 2pt> at 2   0
\put{$x$} [rb] <0pt, 2pt> at -0.4   2.7
\put{$x$} [rb] <0pt, 2pt> at -0.8   3.8
\put{$x$} [rb] <0pt, 2pt> at 0.15  1.3
\put{$\tau$} [lt] <2pt, 0pt> at 0 4
\put{$y$} [t] <0pt, -3pt> at -1 0
\put{$A$} [t] <0pt, -4pt> at -0.75 1.75
\put{$E$} [t] <0pt, -4pt> at 0.4 2.25
\put{$C$} [t] <0pt, -4pt> at 0.85 1.2
\endpicture
\]
 \caption{Classification of  paths with $y\le 0$.
 Dashed: Type A. Solid: Type E. Heavy: Type C.}
 \label{orbitsII}
 \end{figure}

In semiclassical analysis one needs to solve the two-point 
boundary-value problem where $q(0)=y$ and $q(t)=x$ are 
prescribed.  It follows from the foregoing remarks that five 
types of path are possible, which we letter in order of 
increasing complexity and indicate in Figures 
\ref{orbitsI}--\ref{orbitsII}.
 \begin{itemize}
\item For $y > 0$: 
\subitem If $x \geq 0$: Type B 
\subitem If $x<0$: Type D
\item For $y \le 0$: 
\subitem If $x \geq 0 $: Type C
\subitem If $x < 0 $: Type A or  Type E 
 \end{itemize}

As the last condition shows, it is possible for two points in 
space-time to be joined by more than one classical path.  
Conversely, for certain values of $y$, $x$, and $t$ it is 
possible that a path of a certain expected type will not exist.  
The precise constraints on the parameters will depend on 
$\alpha$.  (For example, when $\alpha=2$ a complete excursion 
within the harmonic oscillator potential must take an elapsed 
time of precisely half a period, thus in type~E there is 
a lower bound on~$t$ that does not apply when $\alpha=1$.)

For more refined semiclassical approximations it will be 
necessary to solve the two-point boundary-value problem with $p$, 
$x$, and $t$ given. For each such data list, the sign of $y$ may
not be immediately obvious, so one must explore several of the 
five types for existence of paths.  This problem is left for 
later work.

 All the paths discussed so far are ``zeroth-order'' paths, needed
 to construct the basic WKB propagator $K_\mathrm{scl}\,$.
 The first-order approximation \eref{firstterm} will introduce
concatenations of paths of types C and D, with a possible change of 
velocity at the joint, whose contributions must be 
 integrated over the location of the joint.
Furthermore,
 because our model potentials are not smooth at $x=0$, there is 
another sort of path that must be considered in a complete 
semiclassical treatment.  These are the paths that reflect off the 
singularity at the origin. The contributions of such terms decrease 
rapidly with $\alpha$ (at least for integer~$\alpha$), but for 
$\alpha=1$ (considered in \sref{linear}) they are at least 
comparable to~\eref{firstterm}.  We hope to return to them in later 
work.

\section{The Harmonic Oscillator and the Quadratic Wall} 
\label{quad}

We now consider in detail the model with $\alpha=2$,
\begin{equation} V(x) = 
\cases{
0   & if  $x  < 0$, \cr
\frac{1}{4} \omega^2 x^{2} & if $ x  \geq 0$.
\cr}
\label{quadpot}\end{equation}
 To simplify the formulas we take the mass 
$m$ to be $\frac12$.
As previously noted, each classical path contributes to the
leading-order semiclassical approximation to the propagator.
We calculate the action~$S$, the amplitude~$A$, and the 
Laplacian $\Delta A$.  The Laplacian is of interest because it is 
a crucial factor in the source term  for the next-order 
approximation (see \eref{Q} and \eref{firstterm}).
 In other words, $\Delta A/A$ is the 
residual in the leading-order approximation (the right-hand side 
of \eref{HJquantum}); a singularity in it, in particular,  
signals a breakdown in the approximation.

 Because we are considering a system with time-translation invariance,
as explained in \sref{theory} we usually use the notation
$K(x,y,t -\tau) $
 for the quantum propagator $K(x,t;y,\tau)$.
But we must retain the initial time variable $\tau$ to facilitate 
concatenation of paths.

\subsection{Type A: Free Particle}\label{freespace}
 Whenever  $x$ and $y$ are both negative 
 (i.e., the initial and final points are on the left side of the 
origin,
  Figure \ref{orbitsA}),
there always exists a direct path between them that stays in the 
force-free region.
  The particle's position and velocity then are 
\begin{equation}\label{freepath}
 q(\tau) = y + \tau \bigg (\frac{x-y}{t} \bigg ) , \qquad
    v(\tau) = \dot{ q }(\tau) = \frac{x-y}{t}\,. 
\end{equation} 
 The classical action and 
amplitude of the free direct path are thus
\begin{equation}\label{directS} 
S = \int_{0}^{t} \frac{1}{4} \dot{q}^{2} \, d\tau = 
\frac{(x-y)^{2}}{4t}\,, \qquad A^{2} \equiv - \frac{\partial^{2} 
S}{\partial y \partial x} =\frac{1}{2t} \,.
\end{equation}
 (We note that the Lagrangian for a path of this type is equal to the 
(constant) kinetic energy of the particle.)
The WKB construction \eref{WKBkernel} therefore yields the propagator
\eref{freepropagator} (with $n=1$ and $m=\frac12$).
 The Laplacian $\Delta A$ is $0$, because 
 in this case the WKB propagator is exact:
\begin{equation}\label{propagator1D}
\fl{
K_{f}(x,y;t)=\frac{1}{2\pi}\int_{-\infty}^{+\infty}dk\,e^{ik(x-y)} 
e^{-i\hbar k^2 t/(2m)}=\left(\frac{m}{2\pi i\hbar t}\right)^{1/2} 
e^{-m(x-y)^2/(2i\hbar t)} 
}.
 \end{equation} 
 \begin{figure}
\[
\beginpicture
 \setcoordinatesystem  units <1.5cm, 1.05cm>
 \putrule from -3 0 to 1 0
 \putrule from 0 -0.5 to 0 3
 \put{$\bullet$} at -0.45 2.2
  \put{$\bullet$} at -2.0  0
 \plot -2.0  0
  -0.45 2.2 /
 \put{$x$} [rb] <0pt, 2pt> at -0.481   2.26
 \put{$\tau$} [lt] <2pt, 0pt> at 0 3
 \put{$y$} [t] <0pt, -3pt> at -2.1 0
 \put{$q$} [lt] <2pt, 0pt> at 1 0
\endpicture
\]
 \caption{Type A }
 \label{orbitsA}
 \end{figure}

\subsection{Type B: Harmonic Oscillator}

\begin{figure}
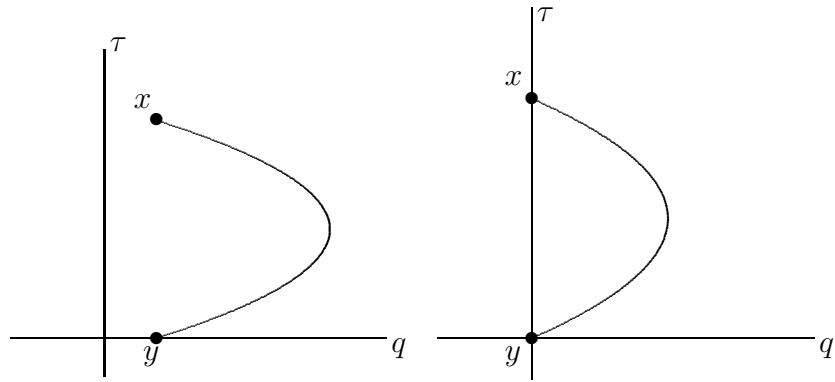

\[
\beginpicture
 \setcoordinatesystem  units <1.25cm, 1cm>
 \putrule from -1 0 to 3 0
 \putrule from 0 -0.5 to 0 3.85
 \put{$\bullet$} at 0.55 0
  \put{$\bullet$} at 0.55  2.9

 \setquadratic
\setsolid\noindent\plot 0.55 0
 2.40 1.45
  0.55 2.9 /  
 \put{$x$} [rb] <0pt, 2pt> at  0.5   3.0
 \put{$\tau$} [lt] <2pt, 0pt> at 0 4
 \put{$y$} [t] <0pt, -3pt> at 0.5 0
 \put{$q$} [lt] <2pt, 0pt> at 3 0
\endpicture
\quad
\beginpicture
 \setcoordinatesystem  units <1.25cm, 1.1cm>
 \putrule from -1 0 to 3 0
 \putrule from 0 -0.5 to 0 4
 
  \setquadratic
\setsolid\noindent\plot 0 0
 1.45 1.45
  0 2.9 /  
  
 \put{$\bullet$} at 0 0
  \put{$\bullet$} at 0  2.9
 \put{$x$} [rb] <0pt, 2pt> at  -0.1   3.0
 \put{$\tau$} [lt] <2pt, 0pt> at 0 4
 \put{$y$} [t] <0pt, -3pt> at -0.2 0
 \put{$q$} [lt] <2pt, 0pt> at 3 0
\endpicture
\]
 \caption{  Left: Type B{}. Right: Special type B}
 \label{orbitsB}

 \end{figure}

 Now let $x$ and $y$ both be in the potential region 
 (i.e., to the right of the origin, Figure \ref{orbitsB}).
Eventually, the boundary cases where one of the coordinates is 
equal to $0$ will be included. 
 
The general solution for a particle that remains inside the 
potential is 
 \begin{equation}\label{quadraticWallClassicalPath} 
q(\tau) = a \cos(\omega \tau) + b \sin(\omega \tau).
  \end{equation} 
Consider the
initial condition 
  \begin{equation}y= q(0) = a  \end{equation}
  and the final condition
\begin{equation} x = q(t) =  y \cos (\omega t) + b \sin(\omega t)  
,\end{equation}
 which implies
\begin{equation} \label{bformula}
  b = \frac{x-y \cos(\omega t)}{\sin(\omega 
t)}. \end{equation} 
 For future use we restate the solution for a general starting time, $s$: 
\begin{equation} \label{oscpath}
  q(\tau) = y \cos(\omega (\tau-s)) +  \frac{x-y 
\cos(\omega (t-s))}{\sin(\omega (t-s))}\sin(\omega (\tau-s)).
  \end{equation} 
 (We always assume $s<t$.)

Returning to $s=0$, we find the
 Lagrangian
  \begin{eqnarray}
L(x,y, \tau)   =& \frac{\omega^{2} }{4} \left [
  \frac{  x^{2} - 2 x y \cos(\omega t ) 
 + y^{2} \cos(2 \omega t )  }{ \sin^{2} ( \omega t)}\,
  \cos(2 \omega \tau) \right. \nonumber\\
&{} - 2 y \left. \left  ( \frac {x - y \cos( \omega t)}{\sin(\omega t) } 
 \right )
\sin ( 2 \omega \tau)    \right ]
\label{Lagrangian1} \end{eqnarray}
 and hence the action
\begin{eqnarray} 
S(x,y,t)&  = \int_{0}^{t} L(x,y, \tau) \, d\tau  \nonumber \\
& = \frac{\omega}{4 
\sin(\omega t)} \big [ x^{2} \cos(\omega t) + y^{2} \cos(\omega 
t) -2 xy ].
\label{SB}\end{eqnarray}
The amplitude is given by 
\begin{equation} A^{2}  = -\, 
\frac{\partial^{2} S}{\partial x \partial y}  
= \frac{ \omega}{2\sin(\omega t)} \,.
\end{equation} 
 As in the free case, $A$ is independent of $x$  and hence $\Delta 
A=0$.
So the WKB propagator for this type  of path is exact and equals
\begin{equation} \label{osckernel}
\fl{ 
K_\mathrm{HO}(x,y,t) = 
 \sqrt[]{ \frac{\omega}{4 \pi i \sin(\omega t)}} \exp 
\biggl \{ - \,\frac{\omega}{4 i \sin(\omega t)} \big [ x^{2} 
\cos(\omega t) + y^{2} \cos(\omega t) -2 xy \bigl ] \biggr \} 
}.
\end{equation} 
 This is the well known \cite{BB97, MF, TT} 
 quantum propagator for the one-dimensional 
harmonic oscillator for $0<t<\pi/\omega$.

 We pause now to note the consequences of the fact that the period 
of a harmonic oscillator is independent of the amplitude.
 A trajectory of the sort shown in Figure \ref{orbitsB}(right),
 starting and ending at the origin, exists only when $t-
s=\pi/\omega$, and in that case it is not unique  (the amplitude 
is arbitrary).
 If either $x$ or $y$ is strictly positive, the (unique) path 
 \eref{oscpath}
 is a segment of one of those paths, and thus necessarily 
 $t-s<\pi/\omega$. The formulas \eref{bformula}--\eref{SB}
  remain meaningful and 
correct when one, but not both, of the endpoints is~$0$.
For example, when $y=0$ 
 the action just simplifies to 
\begin{equation}\label{specialS}
  S(x,0,t) = \frac{\omega x^{2} \cos(\omega t)}{4 
\sin(\omega t)}\,. \end{equation}
 
 In the special case when $x=y=0$ and $t=\pi/\omega$, one must 
return to the general solution \eref{quadraticWallClassicalPath} 
and impose the boundary conditions 
 \begin{equation}
0=y= q(0) , \qquad 0 = x = q(t)
\end{equation}
and a supplementary condition,
\begin{equation}
\dot{q}(0)= v .
\end{equation}
One  finds the solution
 $q(\tau) = b \sin(\omega \tau  ) $
along with the relation
$ b = v/ \omega$ ,
 the consistency condition $t = \pi/\omega$,
 and the final velocity $\dot{q}(t) =  -v$, 
 which is obvious from conservation of energy.
 In terms of a general starting time,
\begin{equation}
q(\tau) = \frac{v}{\omega}\, \sin(\omega (\tau-s)  ) .
\label{specialoscpath} \end{equation}

 In this special case the Lagrangian is 
\begin{equation}
L = \frac{1}{4} \big[ b^{2} \omega^{2} \cos^{2} (\omega \tau) 
- b^{2} \omega^{2} \sin^{2} (\omega \tau) \big],
\end{equation}
and so the action is 
\[ S  = \frac{1}{4} v^{2}  \int_{0}^{t}
% \bigg( \cos^{2} (\omega \tau) - \sin^{2} (\omega \tau)  \bigg)
\cos(2\omega t) \, d\tau  = 
\frac{1}{8\omega} v^{2}  \sin (2\omega t);
\] 
but $ t =   {\pi}/{\omega} $, so
\begin{equation}\label{specialoscaction}
S = 0
\end{equation}
 for these special paths.

 The nonuniqueness of the special paths  and the accompanying 
singularity in~$A$ at $t=\pi/\omega$ can be regarded as a caustic.
 However, it is a very unusual kind of caustic, inasmuch as it does 
not constitute a breakdown of the WKB approximation.
 The function $K_\mathrm{HO}$ is an exact  solution 
of the Schr\"odinger equation when extended through the singular 
point with the correct Maslov phase \cite{TT,MF}.

\subsection{Type E}\label{subsecTypeE}
  The next simplest case to analyze is when
 $x$ and $y$ are both negative  but
  the path goes
through the potential region (Figure \ref{orbitsE}). 
 Such a trajectory is effectively a concatenation
of three types: one part of the path is the special case of
type~B and the other two parts are instances of type~A.  
 Let $t_1$ be
 the time when the particle first crosses into the potential
region and $t_2$ be the time where it exits the potential region.
 Because the potential is quadratic, we know immediately that 
 $t > \frac{\pi}{\omega}$ and  $t_2 = t_1 + \frac{\pi}{\omega}$. 
 Also, 
\begin{equation}
v(t_{1}) = - \,\frac{y}{t_{1}}>0\,
\hbox{ and } \,v(t_{2}) = \frac{x}{t- t_{2}} < 0.
\end{equation}
By conservation of energy, $v(t_{1}) = - v(t_{2})$ and hence
\[
 \frac{y}{t_{1}} = 
\frac{x}{t - 
\biggl (t_{1}+ \frac{\pi}{\omega} \biggr ) }\,. \]
 Therefore, 
 \begin{equation}
  t_{1} = 
\frac{y}{x+y} \biggl (t - \frac{\pi}{\omega} \biggr ) ,
\end{equation}
 and hence
 \begin{equation}
  t_{2} = \frac{ y (t - 
\frac{\pi}{\omega} ) }{x+y} + \frac{\pi}{\omega} \,.
 \end{equation}
The trajectory 
can now be determined from \eref{specialoscpath} and 
 the initial datum
\begin{equation} 
\dot{q}(t_{1}) = - \,\frac{y}{ t_{1} } = - \,\frac{(x+y ) }{t - 
\frac{\pi }{\omega} } \end{equation}
 to be
\begin{equation} q(\tau) = 
\frac{(x + y )  }{ (\pi  - \omega t ) } \sin(\omega (t - t_{1} )).
  \end{equation} 
  \begin{figure}
\[
\beginpicture
 \setcoordinatesystem  units <1.35cm, 0.95cm>
 \putrule from -3 0 to 3 0
 \putrule from 0 -0.5 to 0 4
 \put{$\bullet$} at -2 0
  \put{$\bullet$} at -2.0  3.5
 \plot -2.0  0
  0 0.80 /
 \plot 0 2.9  
       -2.0 3.5 /

 \setquadratic
\setsolid\noindent\plot 0 0.8
 1.80 1.85
  0 2.9 /  
 \put{$x$} [rb] <0pt, 2pt> at -2.1   3.5
 \put{$t$} [lt] <2pt, 0pt> at 0 4
 \put{$y$} [t] <0pt, -3pt> at -2.1 0
 \put{$q$} [lt] <2pt, 0pt> at 3 0
\endpicture
\]
 \caption{Type E }
 \label{orbitsE}
 \end{figure} 

 Now, we compute the 
action by integrating the Lagrangian along the entire path:
\begin{eqnarray}
S_{E}(x,y,t )  &   = S_{A1}(x,y,t) +   S_{B}(x,y,t) +   S_{A2}(x,y,t) 
 \nonumber  \\
& = \int_{0}^{t_{1}} L(q, \dot{q}) \, d\tau + \int_{t_{1} 
}^{t_{2}} L(q, \dot{q}, ) \, d\tau + \int_{t_{2} }^{t_{3}} L(q, 
\dot{q}), d\tau .
\end{eqnarray} 
                But $S_B=0$ according to \eref{specialoscaction},
 and the other two terms are cases of \eref{directS}.
 Thus the total action is
\begin{equation}
  \label{typeEtrajectory} S_{E}(x,y,t) = 
%S_{A1}(x,y,t) + S_{A2}(x,y,t) = 
 \frac{y^{2}}{4 t_{1}} + 
\frac{x^{2}}{4(t -t_{2}) }= \frac{(x+y)^{2}}{4 \big ( t - 
\frac{\pi}{\omega} \big )}\,.
\end{equation} 
The amplitude function is then
\begin{equation} A^{2} = -\, \frac{\partial^{2} S}{\partial x 
\partial y}  =-\, \frac{1}{2 \big (t 
-  \frac{\pi}{\omega} \big )  }\,,
\label{Eamp} \end{equation}
and once again $ \Delta A$ vanishes, 
 Therefore, the solution for type~E is also exact
 (but this string of good luck is about to end). 

 The negative sign in \eref{Eamp} should not be a surprise.
By continuity from the 
(purely harmonic) case ($y = 0$), when $y < 0$ but $y$ is small
  one would 
expect a caustic to occur somewhere near $x = -y$, $t = \pi/ 
\omega$. Therefore, when the trajectory reemerges from the 
potential, this term of the kernel carries  a 
Maslov phase factor of $-i$ \cite{MF,Little,TT}. 
 The existence of the caustic will be verified in the next 
subsection.

\subsection{Type C}  \label{quadC}
\begin{figure}
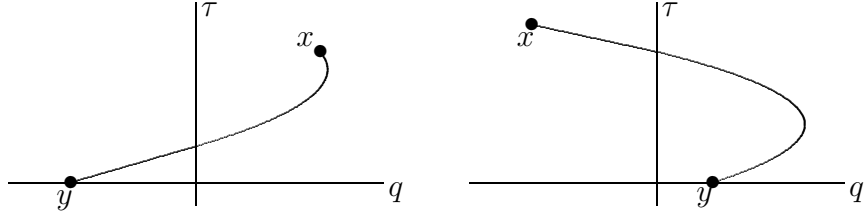

\[
\beginpicture
 \setcoordinatesystem  units <.83cm, 0.60cm>
 \putrule from -3 0 to 3 0
 \putrule from 0 -0.5 to 0 4
 \put{$\bullet$} at 2 2.9
  \put{$\bullet$} at -2  0
 \plot -2.0  0
  0 0.80 /

 \setquadratic
\setsolid\noindent\plot 0 0.8
 1.80 1.85
  2 2.9 /  

 \put{$x$} [rb] <0pt, 2pt> at 1.9   2.9
 \put{$\tau$} [lt] <2pt, 0pt> at 0 4
 \put{$y$} [t] <0pt, -3pt> at -2.1 0
 \put{$q$} [lt] <2pt, 0pt> at 3 0
\endpicture
\qquad
\beginpicture
 \setcoordinatesystem  units <.83cm, 0.60cm>
 \putrule from -3 0 to 3 0
 \putrule from 0 -0.5 to 0 4
 \put{$\bullet$} at 0.9 0
  \put{$\bullet$} at -2  3.5
 \plot 0 2.9  
       -2.0 3.5 /

  \setquadratic
\setsolid\noindent\plot 0.9 0
 2.35 1.46
  0 2.9 /  
 \put{$y$} [rb] <0pt, -3pt> at   0.9   -0.3815 
 \put{$\tau$} [lt] <2pt, 0pt> at 0 4
 \put{$x$} [t] <0pt, -3pt> at -2.1 3.5
 \put{$q$} [lt] <2pt, 0pt> at 3 0
\endpicture
\]
 \caption{Type C (left figure)  and Type  D (right figure)}
 \label{orbitsC}
 \end{figure}

  Here  we have a concatenation of
type A with type B (see Figure \ref{orbitsC}, left). 
  Again  we let $t_1$ denote the
 time  when the particle passes from the free
 region to the  potential region.
  For the first segment of the path, the velocity is
\begin{equation} \label{initv}
\dot{q}(t_{1}) = \dot{q}(0)) =% -\,\frac{y}{t_{1}}\, . 
 -y/t_1\,.
\end{equation}
For the
segment of the path inside the potential, 
 we use the solution \eref{oscpath} found for type~B,
with $y=0$ and  $s=t_1\,$:
\begin{equation}
q(\tau)  = - \,\frac{x}{\sin(\omega(\tau- t_{1})}\,
  \sin (\omega (\tau - t_{1})).
\end{equation}   
 Alternatively, since \eref{specialoscpath}
  with $v$ given in \eref{initv} is independent of~$x$, it applies 
  to our path:
\begin{equation}
q(\tau)  = - \,\frac{y}{\omega t_{1}} \sin (\omega (\tau - t_{1})).
\end{equation}   
 Combining these two equations (or simply setting $x=q(t)$ in the 
second one) yields the relation
 \begin{equation} \label{omegaeq}
 \omega  x t_{1} + y  \sin ( \omega ( t - t_{1} ) ) = 0 
\end{equation}
to determine~$t_1\,$.

 The transcendental equation \eref{omegaeq} for $t_1$ cannot be 
solved in closed form, but
by rescaling the physical variables  to  dimensionless
quantities,  we can get simple master plots of the linear and sinusoidal
terms, as shown in Figure~7.  Intersections
of the lines with the sinusoid represent solutions to \eref{omegaeq}
 and yield valid classical paths.
 Let 
 \begin{equation}
 \Theta= \omega t_{1}\,, \qquad \rho = -y/x, \qquad T= \omega t .
 \end{equation}
  Then \eref{omegaeq} becomes 
\begin{equation} \Theta = \rho \sin (T- \Theta). \end{equation} 
 Setting
 $\Omega = T - \Theta= \omega(t-t_1)$  improves  the equation further to
\begin{equation} T - \Omega = \rho \sin \Omega. \end{equation} 
 Note that
 $t_{1} < t < t_{1} + \frac{\pi}{\omega}$ and hence $\Theta 
< T < \Theta+ \pi$, or $0 < \Omega < \pi$. Thus our task is to find 
zeros, in that interval, of
\begin{equation} f(\Omega) = \rho \sin( \Omega) + \Omega - T.
\end{equation}
 It is clear from the figure that the number of such solutions 
can be  $2$, $1$, or~$0$.
 Observe that if
 \[
 0= f'(\Omega) = \rho \cos(\Omega ) + 1 ,
 \]
 then
\[ \Omega= \cos^{-1} \biggl (- \,\frac{1}{\rho} \biggr ),
 \]
 and that this situation can occur
only if $\rho \geq 1$. 
The condition for the sine curve and the diagonal to be tangent is 
 $f(\Omega) = 0 = f'( \Omega)$, whence
 \begin{eqnarray*}
  0 &= f \biggl (\cos^{-1} 
\biggl (- \,\frac{1}{\rho} \biggr ) \biggr ) = \rho \sin \biggl 
(\cos^{-1} \biggl (- \,\frac{1}{\rho} \biggr )  \biggr ) + 
\cos^{-1} \biggl ( - \,\frac{1}{\rho} \biggr )  - T   \\
& = \rho \sqrt{1 
- \frac{1}{\rho^{2} } } + \cos^{-1} \biggl ( - \,\frac{1}{\rho} 
\biggr )  - T .
\end{eqnarray*} 
 Therefore, we define
  \begin{equation} T_{\ast} = 
\sqrt{\rho^{2} -1 } + \cos^{-1} \biggl ( - \,\frac{1}{\rho} \biggr 
) \qquad (\rho \geq 1) \end{equation} 
 and conclude 
 \goodbreak
\begin{itemize}
  \item $T > T_{\ast} \Rightarrow$ no solutions;
 \item $T = T_{\ast} \Rightarrow$ 1 solution; 
 \item $T < T_{\ast} \Rightarrow$ 2 solutions, if $T \geq \pi$;
  \item $T < \pi \Rightarrow$ 1 solution.
  \end{itemize}

\begin{figure}
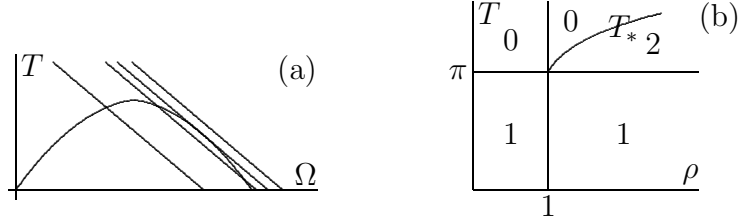

\[ \beginpicture
\setcoordinatesystem units <1cm,1.2cm>
\putrule from -0.1 0 to 4 0
\putrule from 0 -0.1 to 0 1.5
 \put{(a)} [rt] at 4 1.5
 \put{$\Omega$} [rb] <0pt,2pt> at 4 0
 \put{$T$} [lt] <2pt,0pt> at 0 1.5
 \plot   1.355 1.414
         2.355 0.707
         3.355 0 / 
 \plot   1.2 1.414
         2.2 0.707
         3.2 0 / 
 \plot   1.55 1.414
         2.55 0.707
         3.55 0 / 
 \plot   0.5 1.414
         1.5 0.707
         2.5 0 / 
  \setquadratic
 \plot 0 0
 0.785 0.707
 1.57  1
 2.355 0.707
 3.14  0 /
 \endpicture \qquad\qquad
 \beginpicture
\setcoordinatesystem units <1cm,0.5cm>
\putrule from 0 0 to 3 0
\putrule from 0 0 to 0 5
 \putrule from 0 3.14 to 3 3.14
 \putrule from 1 0 to 1 5
 \put{(b)} [lt] at 3 5
 \put{$\rho$} [rb] <0pt,2pt> at 3 0
\put{$T$} [lt] <2pt,0pt> at 0 5
 \put{$0$} at 0.5 4
  \put{$0$} at 1.3 4.5
 \put{$2$} at 2.4 3.9
 \put{$1$} at 0.5 1.5
\put{$1$} at  2.0 1.5
 \setquadratic
 \plot 1 3.14
 1.5 4
 2.5 4.7 /
 \put{$\pi$} [r] <-2pt,0pt> at 0 3.14
 \put{$1$} [t] <0pt, -2pt> at 1 0
 \put{$T_*$} at 2 4.3
 \endpicture\]
 \caption{(a) The four possible relations between a diagonal line and 
the principal arc of the sine curve.
 (b) Resulting division of the parameter plane,
  labeled by intersection numbers.}
 \label{fig:intersects} 
 \end{figure}

  The action for a type C trajectory is the sum of the actions for 
  its two segments,
\begin{eqnarray} %HERE WE DO NEED TWO EQUATION NUMBERS.
S_\mathrm{C}(x, y, t) &= \frac{y^{2}}{4 t_{1}} + 
 \frac{\omega x^{2} \cos(\omega t)}{4 
\sin(\omega t)} \label{SC1}   \\
 &= \frac{y^{2}}{4 t_{1}} + \frac{y^{2}}{8 \omega t_{1}^{2}}
 \sin(2 \omega (t - t_{1} ) ),
\label{SC2} \end{eqnarray}
 where the first version comes from \eref{specialS} and the second follows 
by \eref{omegaeq}. The advantage of the second form is that it is 
independent of~$x$ and hence usable at every point on the curve.

 We turn to the calculation of the amplitude and its Laplacian.
By implicit differentiation  of \eref{omegaeq} in the form
 $y\sin \Omega + x \omega t_{1} =0$
we find
\begin{equation}
\frac{\partial t_{1}}{\partial x} =-\, \frac{t_{1} }{x-y \cos 
\Omega}
\,,
\qquad \frac{\partial t_{1} }{\partial y}
= \frac{x }{y}\, \frac{t_1}{x-y \cos \Omega}\,.
\end{equation} 
Computing   derivatives of \eref{SC2} directly leads to 
complications, but from \eref{SC1} we get the simple formula
 \begin{equation}  \frac{\partial S}{\partial 
y} =  \frac{y}{2 t_{1} } \,.\end{equation}
(This result is recognized as the negative of the initial momentum
  of the particle, as   the Hamilton--Jacobi theory dictates.
It is determined by the initial (free) segment of the trajectory 
regardless of what is considered to be the final endpoint,~$x$;
this explains why only the first term of \eref{SC2} can contribute.)
 We now easily get
 \begin{equation}\label{ampC}
A^{2} = - \,\frac{\partial^{2} S }{\partial x \partial y} =  
  \frac{y}{2 t_{1}^{2} } \,\frac{\partial t_{1} }{\partial x} 
  =-\, \frac{y }{2 t_{1} }\, \frac{1}{x- y \cos  \Omega }\, . 
%  Signs corrected 30 Dec '12.
\end{equation}

There will be a 
caustic if the denominator of \eref{ampC} changes sign. (Since 
that divisor arises from $ \partial t_{1} /\partial x $, its 
vanishing says that $t_{1}$ (hence $y$) can vary without changing 
$x$ (at least to first order).  The caustic thus represents a kind 
of nonuniqueness or degeneracy of the family of paths.)
To study this issue we consider a fixed trajectory with a moving endpoint 
 (that is, fix $y$ and $t_{1}$ and let $x$ and $t$ vary).
  When $t \approx t_{1}\,$, $f'(\Omega)$ is large and positive 
 ($ \rho \to \infty$, $\cos(\Omega) \to 1$). 
 Near the exit point, $t \approx t_{2}\,$, 
 $f'(\Omega)$ is large and negative 
 ($ \rho \to  \infty$, $\cos(\Omega) \to -1$). 
 Therefore, every trajectory does pass through a solution of 
\begin{equation}
0 = f'(\Omega) = \rho \cos (\Omega ) + 1,
\end{equation}
 which is a singularity of \eref{ampC},
somewhere on its retreat from the potential.
 This verifies the last claim in \sref{subsecTypeE}.

In general it is convenient to find the Laplacian of $A$ from $A^2$
  by defining $B = A^{2}$
and noting that
\begin{equation}
\nabla A = \frac{1}{2} B^{-1/2} \nabla B,
\end{equation}
so that %by the chain rule we have
\begin{equation}\label{ABformula}
\frac{\Delta A}A =
  \frac{1}{2} B^{-1} \Delta B - \frac{1}{4} B^{-2} (\nabla B)^{2}.
\end{equation}
In the present case, \eref{ampC} and
\eref{ABformula} yield
 \begin{equation}
 \nabla B = -\,{xy\omega^2 t_1\over 2(x-y\cos(\omega(t-t_1)))^3}
 \end{equation}
 and
\begin{equation} 
\frac{\Delta A}{A }= \frac{ \omega^2 t_{1}^{2}  }{ 4  Y^{4} }
  [4 x y\cos\Omega-6x^2+2y^2 +3x^2\omega^2 t_1^2  ],
\label{Ceffpot}\end{equation}
  where
 \begin{equation}
   Y =  x - y \cos (\omega (t - t_{1}) ) , \qquad
  \Omega = \omega (t - t_{1}) .
 \end{equation}
We observe that $Y$ vanishes at the caustic but the numerator of 
\eref{Ceffpot} does not.
 Therefore, unlike the case \eref{osckernel},
  there is a genuine breakdown
 of the semiclassical propagator approximation
  in the vicinity of the caustic.
 On the far side of the caustic (larger~$t$) the approximation will 
again be good if $A$ (now imaginary) is assigned the phase
  $-i$, in keeping with the general 
Maslov theory \cite{MF,Little}.

\subsection{Type D:  Type C Reversed}
A  type D trajectory has the form
$q(\tau) = a \sin ( \omega \tau  )  + b \cos (\omega \tau )$
inside the potential.
With $q(0) = y$ and $q(t_{1}) = 0$, we arrive at
\begin{equation}
q(\tau) =  -y \cot (\omega t_{1} )   \sin (\omega \tau ) 
 + y  \cos ( \omega \tau) .
\end{equation}
 The total action of the classical particle, in analogy to 
\eref{SC1}, 
 is the sum of that for a B segment and an A segment:
\begin{equation} 
S_{D}(x, y, t)  = 
  \frac { 1 } { 4  }  y^{2}   \omega \cot (  \omega t_{1} )
 +\frac{ x^{2}  }{ 4 ( t - t_{1} )  }\,.
\end{equation}

 Applying the velocity condition to get an implicit equation for
  $t_{1}$ yields
\begin{equation}
\dot{q}(t_{1} ) = \frac{x}{ (t - t_{1}  )  } = 
 - y \omega \cos(\omega t_{1}  ) \cot ( \omega t_{1 }  )
  - y \omega \sin (\omega t_{1}   ),
\end{equation}
which leads quickly to 
\begin{equation} 
y \omega  ( t - t_{1}  ) + x \sin ( \omega t_{1 }  )    = 0 .
\label{implicitD}\end{equation}
 It follows that
\begin{equation}
\frac{  \partial t_{1}   }{ \partial y } = 
 \frac{ t - t_{1}      }{  y - x \cos (\omega t_{1} )  }\,,
 \qquad
\frac{  \partial t_{1}   }{ \partial x }  = 
 -\, \frac yx\, \frac{ t - t_{1}      }{  y - x \cos (\omega t_{1} )  
  }\,.
\end{equation}

 To find  the amplitude function of the trajectory we must
  take the partial derivative of $t_{1}$ with respect to $x$. 
 Because $\frac{\partial S}{\partial x}$ is the final momentum,
 we can obtain the amplitude formula from just the second term of 
$S_D$ if we take the $x$ derivative first:
\begin{eqnarray}
A^{2}   &= - \frac{\partial^{2} S }{\partial y \partial x} 
=   - \frac{\partial  }{\partial y} \biggl [ \frac{ x}
 { 2( t - t_{1})}   \biggr  ] 
 =   -  \frac{ x }{   2 (t - t_{1} )^{2}  } \frac{\partial t_{1} 
}{\partial y}\nonumber\\
& = -\,\frac{     x     }{2 (  t - t_{1}   ) }\,
\frac1{  y- x \cos (\omega   t_1 )      }\,. \label{Damp}
\end{eqnarray}
 From \eref{Damp} we obtain
\begin{equation}
 \nabla B = {y^2\omega^2 (t-t_1)\over
 2(y-x \cos(\omega t_1)^3}\,.
\end{equation}
and hence by \eref{ABformula} 
\begin{equation}
\frac{\Delta A}{A }=-\, \frac{ y^2 \omega^2(t- t_{1})^{2}  }{ 4x^2   X^{4} }
  [4 x y\cos(\omega t_1)-6x^2+2y^2 +y^2\omega^2(t- t_1)^2  ].
\label{Deffpot}\end{equation}
Here $\Omega = \omega ( t - t_{1} )$ and 
 \begin{equation}
 X =y -x \cos (\omega  t_{1} ) .
 \end{equation}
 
 The question of when paths of type D exist is very much like that 
for type C, with the roles of $x$ and $y$ interchanged.
The lack of symmetry between \eref{Ceffpot} and \eref{Deffpot}
arises because the calculation takes place at the endpoint 
 inside the potential in one case but outside in the other.

  In future work we hope to tackle the next
(single-reflection) term, \eref{firstterm},
  in the series \eref{Neumannseries1} by concatenating 
trajectories of types C and D. 
 The final point of the first trajectory is the initial point of 
the second, but the momenta need not match up, since the particle 
is scattering off the effective potential \eref{Ceffpot}.
 For given $(x,y,t)$ outside the potential \eref{quadpot}, one must 
integrate over all $(q,\tau)$ inside the potential for which such 
a path exists. From the taxonomy of paths explained in connection 
with Figure \ref{fig:intersects} 
 it is clear that as many as four trajectories can exist.
Thus the kernel of $K_\mathrm{scl}Q$  turns out to 
be a sum of four terms, each with a domain of integration that is 
a nontrivial subset of the region $(0, \infty) \times (0,t)$.

More precisely, in the C trajectory let us rename $x$ as $q$ and $t$ 
as $\tau$ and add on a new D trajectory from $(q, \tau)$ to $(x,t)$ 
(with $x <0$ and $ t > \tau$).
  The role of $y $ is now played by $x$, and that of $\tau$ 
(which was formerly $t$) is now played by $t-\tau$.
 For the new trajectory we have a new time parameter $\hat{T}$ and
 a new  scaled position parameter $\hat{\rho} 
= - \frac{y}{q}$. 
  There are no solutions if $T\gg 
\max\{\pi, \rho \}$ or $\hat{T} \gg \max\{\pi, \hat{\rho} \}$. If 
$T\gg \frac{\pi}{\omega}$, then $ \rho = - \frac{y}{q}$ and $ 
\hat{\rho} = -\frac{x}{q}$. If $ t\gg \frac{\pi}{\omega} $, these 
conditions 
are both satisfied only for very small $ q $ 
 (large $ \rho $ and~$ \hat{\rho} $). 
 For small $\rho$ there are also solutions with large 
$ q $.
The boundary curves separating the integration domains are 
indicated in Table~\ref{domains}. 
  Further analysis is deferred to future papers.

 \Table{\label{domains}Important boundaries} 
\br&\hfil Old path \hfil &\hfil New path\hfil&\\
 \mr \openup2pt
\hfil$\rho$\hfil  &\hfil $y = - q $\hfil&\hfil    $x = - q$\hfil\\ 
\hfil  $T = \pi $ \hfil& \hfil $\tau = \frac{\pi}{\omega}$\hfil &  
\hfil $t - \tau = \frac{\pi}{\omega}$\hfil \\
 \noalign{\smallskip}
\hfil $ T = T_{\ast}$\hfil &
  \hfil $\tau = \frac{1}{\omega q} \sqrt{y^{2} - q^{2} }
 + \cos^{-1} \big (\frac{q}{y} \big ) $  \hfil
&\hfil $t - \tau = \frac{1}{\omega q} \sqrt{x^{2} - q^{2}} +
 \cos^{-1} \big (\frac{q}{x} \big )$\hfil \\
\br\endtab

\section{The Linear Potential Wall} \label{linear}
 In this section we consider the wall potential with $\alpha=1$:
\begin{equation}
V(x) = 
\cases{
0   & if  $x  < 0$, \cr
k x & if  $x  \geq 0$.
\cr}
\end{equation}
This problem is in some ways harder than that of the quadratic 
potential, because the period of the motion in type B is no longer 
independent of the amplitude.   Also, we will find that the WKB 
construction is no longer exact for type~E.  On the other hand, the 
equations that implicitly determine times when the trajectories 
cross the vertical (time) axis are not trigonometric but cubic and 
hence can be  explicitly solved.

 As usual, we set $ \hbar =1 $ and $ m = \frac{1}{2} $, 
 and the variable $x$ will be replaced by $ q(\tau) $ when we are 
calculating an entire trajectory starting at $y$ and ending at~$x$.
 The equation of motion, inside the potential region 
 (positive $q$-axis),   is 
\begin{equation}
\ddot q(\tau)  = - \frac{k}{m} = -2k ,
\end{equation}
with the general solution
\begin{equation}
q(\tau) = - k  \tau^{2} + a \tau + b.
\label{lingensol}\end{equation}

\subsection{Types A and B} \label{ABlin}
Type A for the linear potential  is, of course,
  exactly the same as before, so the formulas in \sref{freespace}
 still apply.

 For type B we must use \eref{lingensol} with prescribed endpoints,
 \begin{equation}
q(0) = y \ge 0, \qquad q(t) =  x \ge 0.
 \end{equation}
 The results are
 \begin{equation}
   b = y , \qquad a = \frac{(x-y)+ k t^{2}}{t}\,.
\label{linab}\end{equation}
 but it is convenient to refrain from subtituting the cumbersome 
expression for $a$ until after the following calculations are 
finished.
We have
\begin{equation}
\dot{q}(\tau) = -2 k \tau + a  
 \label{linv} \end{equation}
 and hence
 \[
 \dot{q}^{2} = 4 k^{2} \tau^{2} - 4 k \tau a + a^{2} , \quad
V(q)  =  - k^{2} \tau^{2} + k a \tau + k y.
\]
Thus the Langrangian is
\begin{eqnarray} 
L(x,y,\tau) & =  
\frac{1}{4}\dot{q}^{2} - k q 
 \nonumber\\
 &= 2 k^{2} \tau^{2} -2 k \tau a - ky + \frac{1}{4} a^{2} ,
\end{eqnarray} 
and the action  for this path is
\begin{eqnarray}
S(x,y,t) & = 
  \int_{0}^{t} L(q, \dot{q}) \, d\tau  \nonumber\\
 &  = \frac{2 k^{2} t^{3}}{3} - k t^{2} a   +  \frac{a^{2} t}{4} - 
kyt, \nonumber
\end{eqnarray} 
which becomes after a final simplification
\begin{equation}
S 
 = \frac{ (x-y)^{2} }{4 t }  - \frac{(x+ y) k t }{2} -  \frac{k^{2} 
t^{3}}{12}\,.
\end{equation}
 Only the first term gives a nonvanishing contribution to
 \begin{equation}
 A^2 = -\,{\partial^2 S\over \partial y\,\partial x} = 
\frac1{2t}\,,
 \end{equation}
 so the amplitude is the same as for a free particle, and $\Delta 
A=0$.

 It follows that the WKB propagator for this type is 
\begin{equation}
\fl{ 
K_{L}(x,y,t) =  \sqrt[]{ \frac{1}{4 \pi i  t }  } 
\exp \biggl [  \frac{i(x-y)^{2} }{4 t} \biggr ]
  \exp \biggl [- \frac{i(x+y)k t }{2} \biggr ] 
 \exp \biggl [- \frac{i k^{2} t^{3} }{12} \biggr ]
}
\label{linprop}\end{equation}
 and that it is exact.
 Indeed, \eref{linprop}
 is a well known expression for the quantum propagator 
 of a particle in a one-dimensional linear potential
 (e.g., \cite{holstein,CN}).
 
 Unlike in the quadratic problem, no special treatment is needed 
when both $x$ and $y$ equal $0$.
 One simply has
\begin{equation}
S_{B}(0,0,t) = - \,\frac{k^{2}t^{3}}{12}.
\end{equation}
Such a trajectory exists for \emph{any} value of $t$
 (with $a=kt=\dot q(0)$, $b=0$, according to \eref{linab}
 and \eref{linv}), in marked 
contrast to the quadratic case.

\subsection{Type E}
 Type E is simply type B attached to two type-A paths. 
 Therefore, the action is
\begin{eqnarray}
S_{E}(x,y,t) &  = S_{A,1}(x,y,t) + S_{B}(x,y,t) + S_{A,2}(x,y,t) 
 \nonumber  \\ 
 & =  \frac{y^{2}}{4t_{1}} + \frac{x^{2} }{ 4 (t-t_{2} ) } 
 - \frac{k^{2}(t_{2} - t_{1})^{3}}{12}\,.  
 \label{actionIV}\end{eqnarray}

 In the potential region we have a parabolic trajectory of the form
\begin{equation} \label{parabpath}
q(\tau) = - k \tau^{2} + a \tau + c
\end{equation}
 that crosses the axis at $\tau=t_1\,$.
 We need a formula for $t_1$ in terms of the path 
data, $(x,y,t)$.
 Solving \eref{parabpath} for the parameter $ a $ we obtain 
\begin{equation}
  a = \frac{kt_{1}{}\!^{2}  - c}{t_{1} }\,.
\end{equation}
Then the initial-velocity condition at  $t_{1}\,$,
 \begin{equation}
\dot{q}(t_{1}) = v_{1} = -\frac{y}{t_{1}} = - 2 k t_{1} + a ,
 \end{equation}
   yields
\begin{equation} \label{parabcoef}
  a = \frac{ -y + 2 k t_{1}{}\!^{2}}{ t_{1} }\,, \qquad  
 c = y - k t_{1}\!^{2}.
\end{equation} 
 Substituting \eref{parabcoef} back into \eref{parabpath} and solving
 $q(\tau)=0$, we find that the second axis-crossing time is
 \begin{equation}\label{t2one}
t_2=\frac{k t_1{}\!^2 -y}{k t_1}\,.
 \end{equation}
 On the other hand, since $v_{2} = -v_{1} \,$, we have 
\begin{equation}\label{t2two}
   \frac{x}{t-t_{2}  }=\frac{y}{t_{1} } \,.
\end{equation}
Combining \eref{t2one} and \eref{t2two}, we arrive at a quadratic equation 
for $t_1\,$, whose solutions are
\begin{equation}\label{timet1}
t_{1}(x,y,t) = \frac{y}{2  (y+ x)}\{t\pm [ t^{2}
  + 4  (x+y)/k ]^{1/2}\}.
\end{equation}
 Existence requires
 \begin{equation}
 t^2 \ge \frac{4|x+y|}k\,.
\label{Econdition} \end{equation}
The qualitative reason for this condition is that short time 
lapses force $\dot q(t_1)$ to be small for the type-B segment
and large for the type-A segments, and these requirements come 
into conflict if $t$ is too small.
 When the inequality \eref{Econdition} is strict, there are two 
allowed paths, one of low velocity that barely enters the potential 
region and one of high velocity that spends most of its time there.
 (For example, if $x=y=-3kt^2/32$, one solution has $t_1=3t/8$ and
 $t_2=5t/8$ and the other has $t_1=t/8$ and $t_2=7t/8$.)

 \begin{figure}
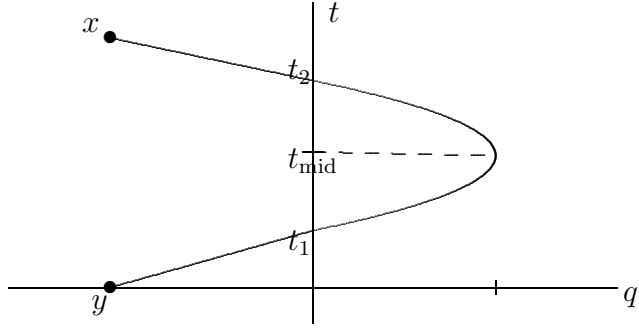

\[
\beginpicture
 \setcoordinatesystem  units <1.35cm, 0.95cm>
 \putrule from -3 0 to 3 0
 \putrule from 0 -0.5 to 0 4
 \put{$\bullet$} at -2 0
  \put{$\bullet$} at -2.0  3.5
 \plot -2.0  0
  0 0.80 /
 \plot 0 2.9  
       -2.0 3.5 /
 \setdashes\noindent\plot  0 1.90 
  1.8 1.85 /
 \setquadratic
\setsolid\noindent\plot 0 0.8
 1.80 1.85
  0 2.9 /  
 \putrule from 1.8 -0.1 to 1.8 0.1
  \putrule from  -0.1  1.9 to 0.1 1.9
 \put{$x$} [rb] <0pt, 2pt> at -2.1   3.5
 \put{$t_{1}$} [lt] <2pt, 0pt> at -0.3 0.8
  \put{$t_\mathrm{mid}$} [lt] <2pt, 0pt> at -0.3 1.95
   \put{$t_{2}$} [lt] <2pt, 0pt> at -0.3 3.2
      \put{$t$} [lt] <2pt, 0pt> at 0.1 4
 \put{$y$} [t] <0pt, -3pt> at -2.1 0
 \put{$q$} [lt] <2pt, 0pt> at 3 0
\endpicture
\]
 \caption{Type E: (Linear Potential Case) The trajectory in the middle
  is not a sinusoidal path in this case, but rather a parabolic one.
  The maximum excursion occurs at time $ t_\mathrm{mid}\, $,
  halfway between the times when the path crosses the axis.}
 \label{orbitsEL}
 \end{figure}

  Substituting  \eref{timet1} and \eref{t2one} into 
  \eref{actionIV},
 we express 
 the action just in terms of the
  $ (x,y,t)$ variables:
\begin{equation}
S_{E}(x,y,t) =  \frac{1}{24} ( k ^{2} t^{2} + 4 k (x+y) )^{3/2}
  - \frac{k t (x+y ) }{4} - \frac{k^{2} t^{3}}{24}\,.
\end{equation}
It follows that 
\begin{equation} \label{ampIV}
A^{2} =    \frac{-\sqrt{k}}{2 \, \sqrt{k t^{2} + 4  (x+y) } }
 \, , \qquad      
{ \Delta A\over A} =\frac{5  }{[ k t^{2} + 4(x+y ) ]^2}
 \,. 
\end{equation}
Clearly the solution is not exact ($\Delta A \neq 0$), and in fact 
the path has passed through a caustic, as shown by the negative 
sign in \eref{ampIV} and the analysis in the next subsection.
 When \eref{Econdition} is an equality, $x$~is actually sitting on 
the caustic point.  
 
 For use in the next subsection we note that if $x=0$ and also the 
caustic equality holds, then
\begin{equation}
t_2{}\!^2 = t^2 = -\,\frac{4y}k\,,\quad t_1=\frac12t_2\,, \quad
t_1{}\!^2  = -\,\frac{y}k\,.
 \label{axiscaustic}\end{equation}

\subsection{Type C}

A path of type C starts at a point $y$ in the free  region and 
ends at a point $x$ in the potential region.  So the classical 
action
  is the sum of the action for the direct path and the action 
 for a type-B path:
\begin{equation}
S_{C}(x,y,t) =   \frac{y^{2}}{4t_{1}} + \frac{x^{2}}{4(t- t_{1})} 
 -\frac{xk(t-t_{1})}{2}  -\frac{k^{2} (t-t_{1})^{3}}{12} \,,
\end{equation}
where $ t $ is the final time and $ t_{1} $ is the time when
  the particle passes through the origin. 

 In the potential region the trajectory has the form 
\eref{parabpath} and must satisfy
\begin{equation}
x=q(t)    = - k t^{2} + a t + c, \qquad
-\, \frac{y}{t_{1} } =\dot{q}(t_{1} )  = - 2 k t_{1} + a  .
\end{equation}
Hence
\begin{equation} \label{acC}
a =2kt_1-\frac y{t_1}\,,\qquad c = x+kt^2-2ktt_1 +\frac{yt}{t_1}\,.
\end{equation}
Therefore, 
\begin{equation}
0=q(t_{1} )
 = +kt_1{}\!^2 -y+x+kt^2-2ktt_1+yt/t_1 \,,
\end{equation}
which  gives a cubic equation for $t_1\,$,
\begin{equation}\label{cubicequation}
 k t_{1}^{3} - 2 k t\, t_{1}^{2} + (k t^{2} + x -y ) t_{1} + yt 
=0.
\end{equation}
With \eref{cubicequation} the formula for $c$ in \eref{acC} can be 
simplified, so that the final formula for the  trajectory is
 \begin{equation} \label{Ctraj}
q(\tau) = -k\tau^2  +\left(2kt_1-\frac y{t_1}\right) \tau
 +(y-kt_1{}\!^2).
 \end{equation} 

 We can rewrite equation \eref{cubicequation} as
\begin{equation}
t_{1}^{3} + a_{2} t_{1}^{2} + a_{1} t_{1} + a_{0} = 0 \label{generalcubic}
\end{equation}
 with
\begin{equation}
a_{2} = -2 t ,  \quad
a_{1} = \frac{x - y}{k} + t^2 , \quad
a_{0}=  \frac{y t}{k}\,.
\end{equation}
A cubic equation always has three roots.  Roots can be either all 
real, with multiplicity 1, 2, or 3,  or one real with the others
 complex conjugates (in the case of real coefficients). 
 The nature of the roots is indicated by a
  polynomial discriminant 
 \cite{Eric,McK,Namias}\cite[p.~17]{NBS}\cite[pp.\ 103--105]{CRC}.
 In terms of  the auxiliary quantities
\begin{equation}
Q = \frac{3 a_{1}  - a_{2}^{2} }{9}\,, \quad 
R = \frac{9 a_{2} a_{1}  - 27 a_{0} - 2 a_{2}^{3} }{54}\,,
\end{equation}
 the cubic equation becomes
 \begin{equation}
 \tau^3+3Q\tau-2R=0, \quad \tau\equiv t-{\textstyle\frac23}t,
 \end{equation}
and the discriminant  is  defined as
\begin{eqnarray}
D & \equiv  R^{2} + Q^{3}\nonumber \\
 &=(3k)^{-3}\left[t^4k^2x +
t^{2}k  \left (  2 x^{2}  + 5 x y - 
{\textstyle\frac14} y^{2} \right ) + (x-y)^3\right].
\label{discriminant} \end{eqnarray}
If $D>0$, one root is real and the other two are complex 
 conjugates. 
  If $D = 0$, all the roots are real with at least two equal.
  And if $D <0$, then all roots are real and distinct. 

   We note that when $t=0$,
\begin{equation}
(3k)^3D =(x-y)^3 >0
\label{D(0)} \end{equation}
(since $y>0$), and also that 
\begin{equation}
\lim_{t \to \pm \infty }  D(t) =  + \infty.
\label{D(infty)}\end{equation}
So, as a function of $t$, either $D(t)$ is everywhere positive or 
it dips negative for some interval of $t$ before becoming positive 
again.
 In the latter case it must have a minimum.
  To analyze this behavior, we note that only positive $t$ 
values are of interest, so we can examine the equation as a 
function of $t^2$ without loss of generality. Letting $T = t^{2} $, 
we see 
 \begin{equation}\label{DT}
 (3k)^3\frac{d D}{d T} =
 2k^2 xT +k\left(2x^2 +5xy-{\textstyle\frac14} y^2\right).
 \end{equation} 
Setting  \eref{DT} to zero  determines the positive critical point
  of $D$ to be  
\begin{equation}\label{Tequation}
  t =  \sqrt{\frac{ -8 x^{2} - 20 x y + y^{2} }{8 x k}}\,.
\end{equation}
 Existence of this critical point requires
\begin{equation} \label{crit}
C\equiv -8 x^{2} - 20 x y + y^{2} \ge 0\,;
\end{equation}
 if \eref{crit} is violated, $D$ is always positive.
 When \eref{crit} holds, 
 substituting \eref{Tequation} into the discriminant 
 \eref{discriminant} gives
\begin{equation}
D_{\min} = -\,\frac{y (8 x+y)^{3} }{1728 x k^3}\,.
\label{critpt}\end{equation}
This expression can have any sign,
  since $x>0$ but $y<0$.
 This shows that all possible cases of cubic roots can occur
 (unlike the related situation in \cite{Todd}, where there are 
always 3 real roots but only the middle one is physically 
relevant).

 Now we define $s=8x+y$, as suggested by \eref{critpt}, and change 
variables from $(y,x)$ to $(s,x)$.   Note that $-\infty<s<8x$.
 From \eref{crit} we have $C= s^2 -36xs+216x^2$.
 Solving \eref{discriminant} as 
a quadratic equation in $T\equiv t^2$, we obtain
 \begin{equation}
T= \frac{27}{4xk}\Bigl({\textstyle\frac14}C \pm \sqrt{P}\Bigr),
\label{Troot} \end{equation}
 where the new discriminant is
 \begin{eqnarray}
 P &\equiv  32 x^3 + 12 x^2y^2 + {\textstyle \frac32} 
xy^3 + {\textstyle \frac1{16}} y^4  \nonumber \\
 &= -{\textstyle \frac12} xs^3 + {\textstyle \frac1{16}}s^4
= {\textstyle\frac{1}{16}} s^3 y.\label{Tdisc}\end{eqnarray}
If $P$ is negative ($0<s<8x$), then there are no real roots 
 for~$T$ and again $D>0$ for all~$T$.
 If $P>0$ ($s<0$), then there are two real roots, and they are 
positive because $C>0$ in this case; thus $D<0$ for some range 
of~$t$.
 It can be seen that $C(s)=0$ has two roots, one slightly less 
than~$8x$ and one in the irrelevant region $s>8x$.
 Thus the results of this analysis can be listed:
 \begin{itemize}
\item $s<0 \imp{}$ 3 real roots exist for some $t$.
 \item $0<s< s_- \approx 7.6x \imp{}$ 1 real root for all $t$.
 \item $s_-<s< 8x \imp{}$ 1 real root for all $t$.
 \item $s>8x \imp y>0$ (forbidden).
 \end{itemize}
 The only distinction between the second and third cases is that in 
the second, $D_{\min}$ is greater than~$0$, while in the third, the 
minimum does not exist at all.

  When there is only one real root, it is conveniently represented 
by the purely algebraic formula developed by
Del Ferro, Tartaglia,  Cardano and Bombelli
  \cite[pp.\ 310--317]{Boyer}:
 \begin{equation}
x=  {\frac23}\,t+
  \root3\of{R + \sqrt{R^2+Q^3}} +\root3\of{R -\sqrt{R^2+Q^3}}\,.
\label{oneroot}
 \end{equation}
  When there are three real roots, however,  formula \eref{oneroot}, 
although still correct, is almost useless in practice.  A better 
formula is the trigonometric representation developed by
  Vi\`ete and  Girard \cite[p.~341]{Boyer}: 
\begin{equation} r_{1}(x,y,t) =  {\frac23}\,t
  + 2\, \sqrt[]{- Q(x,y,t)} \cos 
\biggl ( \frac{\Theta(x,y,t) }{3} \biggr )  , 
 \label{root1}\end{equation} 
\begin{equation} r_{2}(x,y,t) = {\frac23}\,t
  + 2 \, \sqrt{- Q(x,y,t)} \cos 
\biggl ( \frac{\Theta(x,y,t) + 2 \pi }{3} \biggr )  , 
\label{root2}\end{equation} 
  \begin{equation} r_{3}(x,y,t) =  {\frac23}\,t
  + 2 \, \sqrt{- 
Q(x,y,t)} \cos \biggl ( \frac{\Theta(x,y,t) + 4 \pi }{3} \biggr ) ,
 \label{root3} \end{equation}
where 
\begin{equation} 
\Theta(x,y,t)  = \arccos \biggl ( \frac{ R(x,y,t)}{\sqrt{- 
Q(x,y,t)^{3} } } \biggr ). \label{angleTheta}\end{equation} 
 Note that $r_2=r_3$ when $\Theta=0$, which is the same as the 
boundary $D=0$ between the two regimes.

 For some purposes, however, it is convenient to ignore the 
complicated explicit solutions  \eref{oneroot} and 
 \eref{root1}--\eref{angleTheta}
 and work by implicit differentiation as in \sref{quadC}.
 We write 
 \begin{equation}
f(t_1) = k t_{1}^{3} - 2 k t\, t_{1}^{2} + (k t^{2} + x -y ) t_{1} + yt 
\label{cubicfunction} \end{equation}
 for the cubic polynomial in~\eref{cubicequation}.
 In calculating 
$A^{2}$
we shall encounter $\frac{ \partial t_{1}  }{\partial x}$, 
which involves the reciprocal of the derivative of the implicit equation 
$f(t_1)=0$ with respect to $t_{1}\,$. 
A caustic may arise when this factor vanishes. 
As in the case of the quadratic potential,
  a caustic is signaled by a simultaneous solution of 
\begin{equation}
f(t_{1}  ) = 0 , \quad f' (t_{1} ) = 0,
\end{equation} 
i.e., a place where the graph of $f$ is tangent to the horizontal axis. 
 But this is the same as saying that the cubic equation has a 
double root.
 So a caustic occurs at a point where the number of real roots is 
changing from one to three as a parameter of the problem varies.
We define
 \begin{equation}
 Y = f'(t_1) = 3kt_1{}\!^2 -4ktt_1 + kt^2 + x-y.
\end{equation}

 Differentiating \eref{cubicequation}  with respect to 
$x$ yields 
 \[
0 = t_1 +Y \,\frac{\partial t_{1} }{\partial x}\, ,
\]
and thus
 \begin{equation}
\frac{\partial t_{1} }{\partial x}  =
 -\,\frac{t_1}Y\,.
\end{equation}
 Therefore, 
\begin{equation}
A^{2}  = -\, \frac{\partial^{2} S }{\partial x \partial y} =   
\frac{\partial p } {\partial x} = \frac{y}{2 t_{1}^{2} }
\, \frac{\partial t_{1} }{\partial x}
=-\, \frac{y }{2 t_{1}Y }\,.
\end{equation}
Then using \eref{ABformula}, 
 and using the cubic equation repeatedly to reduce the numerator 
 to a quadratic in~$t_1\,$, we obtain
 \begin{equation}
\frac{ \Delta A }{ A  } = \frac{  k }{    Y^{4}   }
 [-24t^2y  -9t(kt^2+2x-y)t_1 +(23kt^2-21x+21y)t_1{}\!^2] .
 \end{equation}

 Finally, we note an interesting special case.
 In the equation $D=0$, set $x=0$ (i.e., consider a caustic 
occurring right on the vertical axis).  Then $t_2=t$, and from 
\eref{discriminant} we have
 \[ 0 =(3k)^3\left( -\, \frac{kt^2y^2}4 - y^3\right),\]
 or
 \begin{equation}
 y = -\, \frac{kt^2}4\,.
\label{cau1} \end{equation}
 But in this situation we also have
 \begin{equation}
t-t_1 = -\, \frac{y}{kt_1}\,,
\label{cau2} \end{equation}
 because the final remark in \sref{ABlin} obviously generalizes to 
the statement that $\dot q(t_1)$ is equal to $k$ times the elapsed 
time of the excursion through the potential region. 
 Substituting \eref{cau2} into \eref{cau1} 
 yields for $y$ the 
equation 
\[ y^2+ 2kt_1{}\!^2+ k^2t_1{}\!^4 =0, \]
 whose only solution is $y=-kt_1{}\!^2$. By comparison with 
\eref{cau1} it follows that  $t=2t_1\,$.
 Thereby we have reproduced the relations \eref{axiscaustic} 
obtained by approaching the axis caustic from the other side.

\subsection{Type $D$}
  The part of a type $D$ trajectory 
in the potential region is determined by the initial conditions 
 $q(0) = y$ and $q(t_{1} ) = 0$. Thus
 \begin{equation} q(\tau) = - k \tau^{2} + a \tau + 
y , \quad
  a= \frac{ k 
t_{1}^{2}  - y}{ t_{1}  } \,. \end{equation} 
  Also, the 
final veolcity  must be the same as the velocity when the
particle crosses the vertical time axis:
\begin{equation} 
 \frac{ x  }{  (t - t_{1}  ) } = \dot{ q }( 
t_{1} ) = - 2 k t_{1}  + a.  \end{equation}
 Hence 
 \begin{equation} a =\frac{ x  }{  (t - t_{1}  ) }  + 2 k 
t_{1}\,. \end{equation} 
 The Lagrangian of the particle in the linear 
potential region is 
 \begin{equation} L(x,y,t) = 
\frac{k^{2} t_{1}^{2}    }{4}  -   \frac{3 k y  }{2}      +    
\frac{ y^{2}  }{4 t_{1}^{2} } -2 k^{2} t_{1}   \tau  +  \frac{2 k y 
\tau }{ t_{1}   }    +    2 k^{2} \tau ^{2}, \end{equation} 
 so the action for the particle in that region is
\begin{equation} S_{B}(x,y,t) = \int_{0}^{t_{1}  }
  L(q, \dot{q},\tau) \, 
d\tau = -\frac{1}{12} k^{2}  t_{1}^{3}  -\frac{k  t_{1} y  }{ 2 }+ 
\frac{   y^{2}   }{4 t_{1}   }\,. \end{equation}
   The total action 
is  the sum of the actions for the two path segments:
\begin{equation} 
 \fl{ S_{D} (x,y, t) = S_{A}(x,y,t) + S_{B}(x,y,t) 
= \frac{   x^{2}  }{  4 (t - t_{1}   ) } -\frac{1}{12} k^{2}  
t_{1}^{3}  -\frac{k  t_{1} y  }{ 2 }+ \frac{   y^{2}   }{4 t_{1}   
} \,.} \end{equation} 
 The final momentum of the classical particle is given 
by $p = - \frac{\partial S}{\partial x}$, and hence
\begin{equation} p =\frac{1}{2} v = \frac{x}{2( t- t_{1} )  } \,.
\end{equation} 
  Taking the partial derivative of momentum with 
respect to $x$ gives
\[  A^{2}  = - \,\frac{\partial^{2} S 
}{\partial y \partial x} =  
 \frac{\partial p } {\partial y} =  \frac{\partial } {\partial y}
  \biggl [   \frac{x}{2( t- t_{1} )  }   \biggr ], 
\]
or 
\begin{equation}\label{amp2}
A^{2} =  \frac{  x }{  2 (  t   -   t_{1}  )^{2}  }
  \frac{\partial t_{1}   }{\partial y}\,.
\end{equation}
 The partial derivative of $t_{1}$ with respect to $x$ can be found 
from 
 \begin{equation}
q(t_{1}  )= 0 = - k t_{1}^{2} + \biggl [ \frac{x}{(t - t_{1}  ) } 
  + 2 k t_{1} \biggr ] t_{1} + y ,
\end{equation}
which simplifies to 
\begin{equation}
0 = k t_{1}^{2} ( t - t_{1} ) + x  t_{1} + y ( t- t_{1} ) .
\end{equation}
 Implicit differentiation  then yields
 \begin{equation}
\frac{\partial t_{1}   }{\partial x}  =  
\frac{t_1}X\,, \quad
\frac{\partial t_{1} } {\partial y} = \frac{t-t_1}X\,,
 \end{equation}
 where
 \begin{equation}
X =   - x  +  y -  (2 k t ) t_{1} +  3 k t_{1}{}\!^{2}.
 \end{equation}
 Thus
 \begin{equation}
A^2 = \frac{x}{2(t-t_1)X}\,.
 \end{equation}
 Finally, after a long \emph{Mathematica} calculation we obtain
 \begin{eqnarray}
&\fl{
 {\Delta A\over A}= \frac1{x^2(t-t_1)^2 X^4}\times{}} \nonumber\\
  &[
t^2y(-4k^2t^4y-47x^2y+76xy^2-20y^3 +6kt^2x^2 -20kt^2xy +40kt^2y^2)
 \nonumber\\
 &{}+2ty(-21x^2+12k^2t^4y +69x^2y -60xy^2+12y^3 -22kt^2x^2 
\nonumber\\
&\hskip2cm{} +60kt^2xy-40kt^2y^2)t_1 \nonumber\\
 &{}+(5x^4+10x^3y+38kt^2x^2 -39x^2y^2 -84kt^2xy^2 +28xy^3 
\nonumber\\
&\hskip2cm{} -20k^2t^4y^2+40 kt^2y^3 -4y^4)t_1{}\!^2] .
 \end{eqnarray}

\section{Conclusion and Outlook} \label{concl}

The WKB, or semiclassical, approximation to the propagator of a 
quantum system in more than one dimension has a well-developed 
literature. Here we have extended it in two directions. 

In \sref{theory} we investigated the convergence of the 
semiclassical series and proved, under rather strong assumptions, 
that it converges, thereby giving a construction of the exact 
quantum propagator as a sum or integral over paths (classical 
except for scatterings off the potential at finitely many 
points).

In later sections we constructed the leading term in the series 
for two instances of a special class of potentials, the soft 
walls of integral power growth.  
Such detailed applications to particular systems are rather rare 
in 
the literature, a fact that becomes less surprising once one 
discovers how complicated the calculations can be even for 
seemingly simple examples.
The first step in such a 
venture is to find all the classical paths between  given 
starting and ending points in space-time; the paths may not be 
unique (and may not exist), and they fall into several 
qualitatively different classes.  For the particular cases of
$\alpha=2$ (harmonic potential inside the wall) and $\alpha=1$ 
(linear potential inside the wall) we studied all the paths in 
detail, constructing their associated semiclassical apparatus of
action, amplitude, and residual function (proportional to the 
Laplacian of the amplitude).  There are noteworthy differences 
between the harmonic and linear cases; generally speaking, the 
former is simpler because of the fixed period of harmonic 
oscillations.

For a solution of the time-dependent Schr\"odinger equation, a 
caustic is not the same thing as a breakdown of the approximation;
  as the propagator 
for the harmonic oscillator shows, a singular function may be an 
exact solution.
For the power walls, however,  it turns out that ``bad''
caustics generically occur; that is, there is a singularity in 
the residual function as well as the amplitude.
Thus the theorem from \sref{theory} ceases to apply after a short 
initial time period.
 
We have left many things undone that are suitable for future 
work.
First, there is the possibility of ``getting over the caustic'' 
by 
constructing a first-order WKB approximation in momentum space.  
We hope that using this improved approximation in the 
higher-order WKB construction will yield a globally convergent 
series.

Of course, we are far from calculating any such series even in 
the caustic-free region.  Our solutions for paths of Types C and 
D are ready to be concatenated to create the first-order term in 
the series by integration over a scattering point inside the 
wall.  Evaluating such integrals in practice may prove 
challenging.

Another problem that has been skipped over here is the 
diffraction from the mild singularity in our potential at the 
origin.  Wherever a potential fails to be of class $C^\infty$,
the semiclassical expansion fails to be valid beyond a certain 
order. A correct approximation in the WKB spirit must then take 
into account exactly the scattering from the point in question.
The resulting correction is small compared to the leading 
semiclassical approximation but will be more significant than 
some higher-order terms in the expansion \eref{Qseries}.
For the power wall with integer exponent~$\alpha$, the 
discontinuity appears in the derivative of order~$\alpha$.
For $\alpha=1$ it is quite significant, and its effects show up 
in the calculation of quantum vacuum energy in that potential 
background~\cite{okla,fuldar}. 

The physical problem motivating \cite{fuldar} and \cite{okla} 
is vacuum energy in relativistic quantum field theory, where the 
natural integral kernel to study is not the Schr\"odinger 
propagator but the Wightman or Feynman wave kernel (or its 
continuation to imaginary time, the cylinder or Poisson kernel).
The latter is not so amenable to WKB techniques, and an 
asymptotic expansion for one kernel does not automatically yield 
one for the other.  However, both kernels encode spectral 
information about the same elliptic operator acting in the 
spatial coordinates, and we continue to investigate how to use 
the nonrelativistic semiclassical construction in the study of 
the relativistic theory.
 
\section*{Acknowledgments} This research has been supported by 
National Science Foundation Grants PHY-0554849 and PHY-0968269.
We thank Jef Wagner for some early involvement and Stuart Dowker 
for a comment.


\Bibliography{10}  \frenchspacing


 \bibitem{NBS} Abramowitz M  and Stegun I A      1964
\emph{Handbook of Mathematical Functions With Formulas,
Graphs, and Mathematical Tables} 
(Washington:U S Dept of Commerce).


\bibitem{Arnold}  Arnold V I 1989 
 \emph{Mathematical Methods of 
Classical Mechanics}, 2nd ed. (Berlin:Springer).
%Corr. 4th printing edition, Springer, December 1, 2010.


\bibitem{BB74}  Balian R and  Bloch C 1974
 Solution of the 
Schr\"{o}dinger equation in terms of classical paths
  \emph{Ann. Phys.} {\bf 85}  514--545.

\bibitem{fuldar}
  Bouas J D,  Fulling S A,   Mera F D,  
Trendafilova C S,  Thapa K and Wagner J 2012
  Investigating the spectral 
geometry of a soft wall, in \emph{Spectral Geometry},
  edited by A H Barnett, C S Gordon, P A Perry and A Uribe, 
 \emph{Proceedings of Symposia in Pure Mathematics} {\bf 84}, pp.\ 
139--154.

\bibitem{Boyer}  Boyer C B 1968
 \emph{A History of Mathematics}
 (Princeton:Princeton U~P).

\bibitem{BB97}
Brack M and Bhaduri R K 1997 
 \emph{Semiclassical Physics}
(New York:Addison--Wesley).

 \bibitem{CN}Carlitz R D and Nicole D A 1985
Classical paths and quantum mechanics
\emph{Ann.   Phys.} {\bf164} 411--462.

\bibitem{dowkw} Dowker J S 1975 Covariant Schr\"odinger 
equations, in \emph{Functional Integration and its Applications}, 
edited by A. M. Arthurs (proceedings of international conference 
at Windsor, 1974), pp.\ 34--52.


\bibitem{Evans} Evans L C 1998
 \emph{Partial Differential Equations},  Graduate Studies in
Mathematics {\bf 19}
(Providence:American Mathematical Society).


\bibitem{folland} Folland G B  1995
  \emph{Introduction to 
Partial Differential Equations}, 2nd ed.
(Princeton:Princeton U~P).

 \bibitem{HJ} Hagedorn G  and Joye A  1999
 Semiclassical 
dynamics with exponentially small error estimates
\emph{Commun. Math. Phys.} {\bf 207}  439--465.


 \bibitem{holstein} Holstein B R  1997
 The linear potental propagator 
 \emph{Amer. J. Phys.} {\bf65}  414--418.

 \bibitem{kress}   Kress R 1999
  \emph{Linear Integral Equations}, 2nd ed., 
 (New York:Springer).




\bibitem{Little} Littlejohn  R G 1992
The Van Vleck formula, Maslov theory, and phase space geometry 
  \emph{J.  Stat. Phys.} {\bf 68} 7--50.

\bibitem{MF}  Maslov V P and  Fedoriuk M V 1981
\emph{Semi-Classical Approximation in Quantum Mechanics}
 (London:Reidel).

\bibitem{McK} McKelvey J P 1984
 Simple transcendental expressions for the roots of cubic equations
 \emph{Amer. J. Phys.} {\bf52} 269--270.

\bibitem{Mera}  Mera F D 2011
\emph{The Schr\"odinger Equation  as a Volterra Problem},
M.S. thesis, Texas A\&M University, 
{\tt 
http://hdl.handle.net/1969.1/EDT-TAMU-2011-05-9194}.


\bibitem{Mera2}  Mera F D and Fulling S A
  2013
Convergence of the Neumann series for the 
Schr\"odinger equation 
and general Volterra equations in Banach spaces,
in \emph{Quantum Mechanics}, edited by P. Bracken 
(Rijeka, Croatia:InTech), in press.

 


\bibitem{okla}  Milton K A 2011
Hard and soft walls
\emph{Phys. Rev. D} {\bf 84}  065028.


 \bibitem{Namias}  Namias V 1985
  Simple derivation of the roots of a cubic equation
 \emph{Amer. J. Phys.} {\bf53} 775. % one page

 \bibitem{putrov}  Putrov P A 2008
Path integral in the energy representation in quantum 
mechanics
  \emph{Theor. Math. Phys.} {\bf156}  1041--1057
 [\emph{Teor. Math. Fiz.} {\bf156}  92--111].


 \bibitem{rub}   Rubinstein I and L. Rubinstein  L    1998
\emph{Partial Differential Equations in Classical Mathematical
Physics}  (New York:Cambridge University Press).


 \bibitem{CRC} Selby S M 1973
 \emph{CRC Standard Mathematical Tables}, 21st ed.\
 (Cleveland:Chemical Rubber Co.).


\bibitem{TT} Thornber N S and Taylor E F 1998
Propagator for the simple harmonic oscillator
 \emph{Amer. J. Phys.} {\bf66} 1022--1024.


 
 \bibitem{Eric}  Wolfram MathWorld 2004
Cubic formula  
{\tt http://mathworld.wolfram.com/CubicFormula.html}.


 
 \bibitem{Todd}  Zapata T A 2007
\emph{The WKB Approximation for a Linear Potential and Ceiling},
  M.S. thesis, Texas A\&M University {\tt 
http://repository.tamu.edu/handle/1969.1/85876}.

   \endbib
 \end{document}